\documentclass[smallextended]{svjour3}       
\smartqed  
\usepackage{graphicx}
\usepackage{cite}
\usepackage{amsmath,amssymb,amsfonts}
\usepackage{algorithmic}
\usepackage{graphicx}
\usepackage{textcomp}
\usepackage{xcolor}
\usepackage{colortbl}
\usepackage{tikz}

\usepackage{amsmath,amssymb,amsfonts}
\usepackage{algorithmic}
\usepackage{graphicx}
\usepackage{textcomp}
\usepackage{xcolor}
\usepackage{longtable}
\usepackage{float}
\usepackage{graphicx}
\usepackage{multirow}
\usepackage{balance}
\usepackage{flushend}
\usepackage{hyperref}
\usepackage{tabularx} 
\usepackage{booktabs} 
\usepackage{hhline}
\usepackage{enumitem}
\usepackage{array}
\usepackage{lipsum}
\newcolumntype{L}[1]{>{\raggedright\let\newline\\\arraybackslash\hspace{0pt}}m{#1}}
\newcolumntype{C}[1]{>{\centering\let\newline\\\arraybackslash\hspace{0pt}}m{#1}}
\newcolumntype{R}[1]{>{\raggedleft\let\newline\\\arraybackslash\hspace{0pt}}m{#1}}
\usepackage{subfigure}
\usepackage{lscape}
\usepackage[
  separate-uncertainty = true,
  multi-part-units = repeat
]{siunitx}

\newcommand{\etal}{ \emph{et} al.}

\newcommand{\code}[1]{ {\tt #1}}
\definecolor{blue_}{HTML}{B8EBFF}
\definecolor{pink_}{HTML}{FFCCCB}
\definecolor{umber}{HTML}{CC0000}
\definecolor{orange}{HTML}{CC5500}
\definecolor{redd}{HTML}{F26035} 

\newcommand{\bluebg}[1]{\colorbox{blue_}{#1}}
\newcommand{\pinkbg}[1]{\colorbox{pink_}{#1}}
\newcommand{\graybg}[1]{\colorbox{gray!25}{#1}}

\newcommand{\revised}[1]{\textcolor{black}{#1}}

\newcommand{\biasToMen}[1]{\cellcolor{blue_}{#1}}
\newcommand{\biasToWomen}[1]{\cellcolor{pink_}{#1}}
\newcommand{\biasToGendered}[1]{\cellcolor{gray!25}{#1}}

\everymath{\scriptstyle}
\journalname{Empirical Software Engineering}

\newenvironment{boxedtext}
    {
    
    \begin{center}

    \begin{tabular}{|p{0.96\linewidth}|}
    \hline
    }
    { 
    \\ \hline
    \end{tabular} 
    
    \end{center}
       }

\def\BibTeX{{\rm B\kern-.05em{\sc i\kern-.025em b}\kern-.08em
    T\kern-.1667em\lower.7ex\hbox{E}\kern-.125emX}}
\begin{document}

\title{Code Reviews in  Open Source Projects : How Do Gender Biases Affect Participation and Outcomes?
 }
 
\titlerunning{How Do Gender Biases Affect Code Review Participation and Outcomes?}
\authorrunning{Sultana \textit{et} al.}

\author{Sayma Sultana         \and
        Asif Kamal Turzo  
        Amiangshu Bosu
}


\institute{S. Sultana, A. Turzo, and A. Bosu \at
              Department of Computer Science, Wayne State University, Detroit, Michigan. \\
              \email {sayma@wayne.edu},  {asifkamal@wayne.edu}, {abosu@wayne.edu},           
          }

\date{Revised: February 7, 2023 / Accepted: TBD}

\maketitle

\begin{abstract}

\textit{Context:} Contemporary software development organizations lack diversity and the ratios of women in Free and open-source software (FOSS) communities are even lower than the industry average. Although the results of recent studies hint at the existence of biases against women, it is unclear to what extent such biases influence the outcomes of various software development tasks.

\textit{Aim:} This study conceptually replicates two recent studies by Terrell \textit{et} al. and Bosu and Sultana that investigated gender biases in FOSS communities. We aim  \textit{ to identify whether the outcomes of or participation in code reviews (or pull requests) are influenced by the gender of a developer.} In particular, we focus on two outcome aspects (i.e., code acceptance, and review interval) and one participation aspect (i.e., code review participation) of code review processes. 
\textit{Approach:} With this goal, this study includes a total of 1010 FOSS projects. Ten out of those projects use Gerrit-based code reviews. The remaining 1000 are randomly selected from the GHTorrrent dataset based on a stratified sampling of projects fitting certain criteria.  We divided GitHub projects into four groups based on the number of distinct contributors. We developed six regression models for each of the 14 datasets (i.e., 10 Gerrit based and four GitHub) to identify if code acceptance, review intervals, and code review participation differ based on the gender and gender-neutral profile of a developer. 

\emph{Key findings:}
Our results find significant gender biases during code acceptance among 13 out of the 14 datasets, with seven favoring men and the remaining six favoring women. We found significant differences between men and women in terms of code review intervals, with women encountering longer delays in three cases and the opposite in seven. 
Our results indicate reviewer selection as one of the most gender-biased aspects with  12 out of 14 datasets exhibiting biases.  A total of 11 out of the 14 cases show that women have significantly lower code review participation than their men peers.  Since most of the review assignments are based on invitations, this result suggests possible affinity biases among the developers.
We also noticed a significantly higher likelihood of women using gender-neutral profiles.  Supporting Terrell \textit{et} al.'s claim, women with gender-neutral profiles had higher odds of code acceptance among three Gerrit-based projects. However, contradicting their results, we found significantly lower odds of code acceptance for women with gender-neutral profiles across all four GitHub project groups. 

\emph{Conclusion:}
 Though gender bias exists among many projects, the direction and amplitude of bias vary based on project size, community, and culture.  Similar bias mitigation strategies may not work across all communities, as characteristics of biases and their underlying causes differ. As women are less likely to be invited for reviews, FOSS projects should take initiatives to ensure equitable selection of women as reviewers.

\end{abstract}

\maketitle

\section{Introduction}
\label{sec:intro}
According to the US labor market census in 2019, women make up approximately 18.8\% of technical roles in the software industry and 18\% of computer science graduates~\cite{labor_market}. 
However, the number of females contributing to Free and Open Source Software (FOSS) projects is even lower (i.e., less than 10\%)~\cite{terrell_1,vasilescu2015gender}; which indicates a significant gender imbalance within the FOSS communities.  Due to the lack of gender diversity, many women hesitate to join FOSS projects, as one quotes, \textit{``I feel like it’s a circle I can’t get into. But mostly I fear the excessive spotlight of being the sole female programmer on a publicly available project. In light of how women are treated on the internet, this fear does not seem unreasonable''}~\cite{open_source_women}. 
The lack of diversity among FOSS communities has been subjected to several prior studies~\cite{oss_floss, floss_pols,DAVID2008364,bosu-esem-2019}.  To increase diversity through recruitment of the underrepresented groups, FOSS communities such as Mozilla~\cite{mozilla-diversity}, Debian~\cite{debian-diversity}, and Fedora~\cite{fedora-diversity}, as well as commercial software development organizations such as Google\footnote{\url{https://diversity.google/}} and Deloitte~\cite{div_deloitte}, have adopted several initiatives. 

However, these diversity initiatives may not achieve adequate results if a community does not promote `inclusion' by providing equal access to opportunities and resources for people who might otherwise be excluded or marginalized~\cite{div_inclusion}. Vernā Myers, a leading diversity and inclusion strategist, says, \textit{“While diversity means being invited to the party, inclusion means to be asked for a dance”}\cite{verna_}. Sense of inclusion drives people to be more productive and engaged with the workplace as well as improves overall team performance and creativity~\cite{verna_,div_inclusion}. Organizations that practice inclusion generate up to 30\% higher revenue per employee and greater profitability than their competitors \cite{div_deloitte}. 

Despite these known benefits, increasing Diversity Equity Inclusion (DEI) among FOSS communities is difficult due to various explicit~\cite{floss-wiki} and implicit biases~\cite{Beatrice18} against minorities. For example, a 2017 study by Terrell \textit{et} al.~\cite{terrell_1} found that women's pull-requests are less likely to be accepted than those created by men if their gender is visibly identifiable. On the other hand, if a woman's gender is not identifiable, her pull requests have higher odds of acceptance than those created by men.
Another study by Bosu \textit{et} al.~\cite{bosu-esem-2019} investigated the state of DEI among ten popular FOSS projects. Their results hint at possible biases against woman developers in three out of the ten cases as women had lower code acceptance rates as well as longer delays than men in getting feedback for submitted changes. Additionally in one project women received significantly lower code review invitations than their men peers.

However, based on the results of this two studies~\cite {bosu-esem-2019,terrell_1}, we cannot predict to what extent developers' genders influence those outcomes,  since several other confounding factors may have been influential as well. For example, the acceptance of a pull request may be influenced by its size, complexity, and whether it includes a new feature or a bug fix. Bosu \textit{et} al.'s study~\cite{bosu-esem-2019} did not account for any of these confounding factors. Although Terrell \textit{et} al.~\cite{terrell_1} conducted post hoc analysis to identify the influences of other factors such as experience, patch size, patch type, and status within the community, it is unclear how much variances in acceptance rates are influenced by the genders of the contributors. To answer these questions, this study aims to conduct a partial conceptual replication~\cite{shull2008role} of these two studies (i.e. Terrell \textit{et} al.~\cite{terrell_1}, and Bosu \textit{et} al.~\cite{bosu-esem-2019}) by adopting measures to control for various confounding factors.

In this replication, we aim \textit{ to identify whether the outcomes of or participation in code reviews (or pull requests) are influenced by the gender of a developer.} However, since the identification of non-binary genders is not feasible without inputs from the subjects, we focus primarily on men and women. 
With this goal, this study includes a total of 1010 FOSS projects. Ten out of those projects are selected to replicate Bosu \textit{et} al.'s study. The remaining 1000 are randomly selected from GHTorrent~\cite{Gousi13} dataset based on a stratified sampling of projects fitting certain criteria.  We divided GitHub projects into four groups based on the number of distinct contributors. 
Similar to Bosu \textit{et} al., we computed genders initially using the GenderComputer tool~\cite{gender_v}, then followed up with manual validations using multiple sources including avatars, and social media profiles. 

We select two outcome aspects and one participation aspect of code reviews: i) acceptance rate-- the ratios of submitted code reviews (or pull requests) that are accepted,  2) review interval -- the time from the beginning to the end of the review process, and  3) code review participation-- the number of code reviews within a period as a reviewer. We developed six regression models for each of the 14 datasets (i.e., 10 Gerrit based and four GitHub), where one of those three aspects is the dependent variable and in addition to the genders of the contributors, various confounding factors that have been found to be influential in patch acceptance by prior studies, are independent variables. We used \textit{multivariate regression modeling} techniques since those can be used to estimate the relationships between a dependent variable and one or more independent variables (aka `predictors') and are commonly used to eliminate effects of confounding variables~\cite{pourhoseingholi2012control,harrell2015regression}. 

\revised{We proposed this study in a registered report titled “Are Code Review Processes Influenced by the Genders of the Participants?” This registered report included our proposed study protocol and was published in  ICSME 2021~\cite{RR_ICSEME}. According to the plan, we executed our study where we considered thirteen characteristics of contributors and pull requests to identify the impact of gender in three aspects: pull request acceptance, code review interval, and code review participation. We have tried our best to execute the study exactly according to our proposed protocol. However, during the study execution, we decided to mine the selected projects again to include the most recent code reviews /pull requests.  We also altered our protocol to determine whether a profile is gender neutral since during execution, we identified several shortcomings of our initial choice (i.e.,  Terrell et al. 's approach).  We have described those deviations in Section ~\ref{profile-comp} and  in a Disclaimer  at the end of this paper.}

In summary, the primary contributions of this study:
\begin{itemize}
    \item Empirical evidence regarding how gender influences code review processes.
    \item An investigation of how gender biases manifest across various project groups.
    \item Recommendations for projects to increase diversity and inclusion.
    \item An update of the status quo of gender diversity among FOSS projects.
    \item To promote replication, we have made our scripts, and dataset (without personally identifiable information) publicly on Zenodo~\cite{replication}.
\end{itemize}
   
The remainder of the paper is organized as follows.
Section \ref{sec:original-studies} briefly describes our replication protocol and the two studies being replicated.
Section \ref{sec:hypotheses} presents formulated hypotheses based on our research questions.
Section \ref{sec:dataset} details our research methodology. 
Section \ref{sec:result} and Section \ref{sec:discussion} presents the results and discusses the implications respectively. 
Section ~\ref{sec:related-works} describes prior related works.
Finally,  Section  \ref{sec:threats} and \ref{conclusion} discusses threats to validity and concludes the paper respectively.

\section{Replication Protocol}
\label{sec:original-studies}
Replications are crucial in the Software Engineering (SE) domain to build knowledge through a family of experiments~\cite{santos}. According to Shull \textit{et} al.~\cite{shull2008role},  SE replications falls into two categories:
\begin{enumerate}
    \item  \textit{Exact replication:} In an exact replication,  study procedures are closely followed. Exact replications can be further divided into two subcategories: i) \textit{dependent} and ii) \textit{independent}. In a \textit{dependent replication},  all the variables and conditions are kept as close to the original studies as possible. However, some of the aspects of the original study can be modified in an \textit{independent replication}  to fit a new context.
    \item  \textit{Conceptual replication:} This category of replications uses different research methodologies to study the same set of questions.
\end{enumerate}
 
Our replications for this study fall into the `partial conceptual replication' category, since-- i)  we plan to use a different type of statistical modeling technique, ii) we will measure and account for additional confounding variables not considered in the original studies, iii) we will use updated versions of the datasets to include recent code reviews to reflect contemporary state of the projects, and iv) we are interested in partial replications to retest only the results from the original studies hinting explicit gender biases.  
The following subsections, briefly describe those two studies.

\subsection{Gender differences and biases during pull request acceptance~\cite{terrell_1}}
To investigate gender biases among FOSS projects, Terrell \textit{et} al. investigated pull request (PR) acceptance on GitHub. The stated goals of their study were \emph{to what extent do gender biases exist among contributors judging PRs for acceptance.} 

\vspace{4pt}
\noindent \emph{Study Design:} Terrell \textit{et} al. used the GHTorrent dataset~\cite{Gousi13}  from June 07, 2010, to April 1, 2015. To estimate the genders of the contributors, they first used the GenderComputer tool~\cite{gender_v}. They augmented this resolution with Google+ profile search based on a contributor's GitHub-listed email address. 
To identify whether a profile is gender neutral or not, they downloaded GitHub profile pictures and used ImageMagick to identify `identicons'\footnote{ 5×5 pixel sprites that are generated using a hash of the user's ID.} They classified a GitHub account as gender neutral if:  i) the profile uses an identicon ii) the gender inference tool outputs `unknown' for both user login name and full name, and iii) their mixed-culture panel judges could not resolve the gender. They considered each pull request as a data point with men forming the `control group' and women as the `treatment'. In addition to gender, they investigated the following confounding factors during posthoc analyses: i) number of lines added, ii) number of lines removed, iii) number of commits, iv) number of files changed, v) pull index (i.e., the author's nth pull request), vi) number of references to issues, vii) licenses (whether a project is open-source or not), viii) creator type (i.e., insider or outsider),  and ix) file extension.

\vspace{4pt}
\noindent \emph{Key results:} The key findings of their study are:

\begin{itemize}[leftmargin=*]
\itemsep0pt
     \item  \emph{Acceptance rate:}  Although they hypothesized women have a lower acceptance rate than men overall, they found support for a contrasting hypothesis. Overall women had 78.7\% acceptance rate, while men had 74.6\% ($\chi^2 =1170, p<0.001$).
  
     \item \emph{Survivorship Bias:} They hypothesized that the acceptance rate starts low for women and surviving women make the majority of pull requests. Their investigation of PRs over time (i.e.,  average acceptance on the first PR, second PR, and so on) found women having higher acceptance rates all along.
     \item \emph{Focus on limited projects:} Their second hypothesis stated that women limit their contribution to fewer projects and maintain higher success rates. While women do tend to work on fewer projects, due to limited data points, their investigation was not conclusive to draw any clear conclusion regarding whether women working on a higher number of projects have higher or lower acceptance rates.
     \item \emph{Pull requests for much-needed features or issues:}  To evaluate if women's PRs fulfill any immediate need, they examined descriptions of PRs and the percentage of requests that refers to any issue. Results showed that women were less likely to submit PRs satisfying immediate project needs.
     \item \emph{Size of changes:} They investigated the number of women's contributions using three metrics: lines of code changed, number of modified files, and number of commits in each pull request. They found out that on average women make larger PRs than men. 
     \item \emph{Success in contributing program code:}  Researchers considered Turing-complete programming languages as program code and showed that women possess a higher acceptance rate in submitting programming codes.
     \item \emph{Appearance as woman:} To find out the effect of being revealed as a woman, the authors divided their analysis into two parts. For insiders, they found out that women possess almost an equal acceptance rate when their gender is identifiable or not. As outsiders, women with gender-identifiable profiles have 10.2\% lower acceptance rates than women with gender-neutral profiles.
 \end{itemize}
 
\vspace{4pt}
 \noindent \emph{Artifacts:} Among the 4.03 million GitHub users, who publicly listed their email addresses, their Google+ assisted linking strategy helped identify genders of  1.4 million (35.3\%) contributors. Since their gender-linking approach may raise privacy concerns, they did not publish their dataset publicly. However, their scripts are publicly available for replication.

\subsection{Diversity and inclusion in FOSS projects~\cite{bosu-esem-2019}}
Bosu \textit{et} al.~\cite{bosu-esem-2019}  investigated  the state of diversity and inclusion among ten FOSS projects based on the following four research questions:
\begin{itemize}
\item \emph{RQ1}: What are the percentages of female contributors in popular FOSS projects?

\item \emph{RQ2}: What percentage of leadership roles are occupied by females among the FOSS projects?

\item \emph{RQ3:} Are female developers of FOSS projects significantly more/less productive than their male colleagues?

\item \emph{RQ4:} Is there any explicit bias for or against female developers in FOSS projects?
\end{itemize}

\noindent \emph{Study Design:} Using a tool to mine Gerrit-based code reviews, they mined data from  ten FOSS projects: i) Android, ii) Chromium OS, ii) Couchbase, iv) Go, v) LibreOffice, vi) OmapZoom, vii) oVirt, viii) Qt, ix) Typo3, and x) Whamcloud. In the first step, they used the GenderComputer tool to resolve the gender of the users. For all non-male users classified by the GenderComputer tool, they executed a five-step manual verification process where they used a step only if its preceding steps were failures. These steps are: i) checking Gerrit's avatar, ii) using Google+, iii) using LinkedIn profiles, iv) using Facebook accounts, and v) using Google search. Using this strategy, they were able to resolve genders for 99\% of the non-casual developers (i.e., developers who had made at least five commits). To investigate the state of diversity and inclusion, they calculated the percentage of non-casual and core developers in those projects. They used commit count-based heuristics to identify core developers, where the group of contributors belonging in the top 10\% based on commit count for a project is the core. To compare the productivity between men and women, the authors considered two metrics: code churn per month and code changes per month, and calculated the median of those. Furthermore, to find out if women encounter explicit biases, authors took into account the following four metrics: i) acceptance rate, ii) first feedback interval, iii) review interval, and iv) code churn per comment.

\vspace{4pt}
\noindent  \emph{Results:} Following points are the result of their study that address the proposed research questions respectively:
 \begin{itemize}[leftmargin=*]
 \item \emph{Gender diversity:} They found that in all of the projects women comprise less than 10\% of the non-casual developers. 
 
  \item \emph{Gender inclusion:} Among six out of the ten projects, the percentage of women as core developers is even lower than the percentage of women as non-casual developers. On average, 4.14\% women among the core developers also showed a poor representation of women in leadership roles.
  
  \item \emph{Productivity comparison:} None of the projects showed any significant differences between men and women in terms of two productivity metrics: i) code churn per month, ii) code changes per month. 
  
  \item \emph{Explicit bias for/against women:} Their results suggest potential explicit biases against women in three among ten projects. In these projects, women had a lower code acceptance rate, had to wait significantly longer to get their initial feedback from the reviewers, and also had to wait a significantly longer time to get their complete review. On the contrary, in three other projects, the result shows biases favoring women. They also showed that men participate more in code reviews. As code review is done based on invitations, authors hypothesized that women may be excluded to favor men. 
 
 \end{itemize}

\noindent \emph{Artifacts:} For all ten projects, authors mined a total of 683,865 pull requests and identified 4,543 developers' information in total. Among the developers, they were successful in inferring 4,496 (99\%) developers' gender. They also did not publish their data publicly due to privacy concerns. However, we do have access to their scripts and dataset due to an overlapping author.

\section{Hypotheses}
\label{sec:hypotheses}
 Our primary goal is to investigate whether the outcomes of or participation in code reviews are influenced by the gender of a developer. 
  Although Terrell \textit{et} al~\cite{terrell_1} found higher acceptance rates for women on GitHub,  Bosu \textit{et} al.'s ~\cite{bosu-esem-2019} results support that finding in only two out of the ten projects and three projects from their study suggested the contrary. Therefore, our first research question is:
 \begin{boxedtext}
\textbf{RQ1:} \emph{
 Do the genders of the contributors influence acceptance of their code changes?}
\end{boxedtext}

While we acknowledge that there are several possible choices for the gender of a person\footnote{\url{https://www.healthline.com/health/different-genders}}, due to the lack of a reliable automated mechanism to identify genders that do not fall into binary choices (e.g. man, woman), prior Software Engineering (SE) studies~\cite{vasilescu2015gender,terrell_1,bosu-esem-2019,lin2016recognizing} confined to comparisons between men and women while studying the impact of gender. Therefore, we plan to limit our investigations to identify the differences between men and women by formalizing this research question into the following two hypotheses:

\boldmath{$H1.1_0$:} \textit{There are no significant differences in code acceptance rates between men and women.}

\boldmath{$H1.1_a$:} \textit{There are significant differences in code acceptance rates between men and women.}

Since  Terrell \textit{et} al. ~\cite{terrell_1} found that women have a significantly lower code acceptance rate than men when their gender is identifiable, we pose two additional hypotheses.

\boldmath{$H1.2_0$:} \textit{There are no significant differences in code acceptance rates between men/women with and without gender identifiable profiles.}

\boldmath{$H1.2_a$:} \textit{There are significant differences in code acceptance rates between men/women with and without gender-identifiable profiles.}

 \vspace{4pt}

`Code review interval' spans from the beginning to the end of the code review process~\cite{def_interval}. Bosu \textit{et} al. ~\cite{bosu-esem-2019} found that women developers had to wait three times longer than men in two out of the ten projects. However, two of their projects suggested the opposite. Our next research question aims to investigate those differences.

\begin{boxedtext}
\textbf{RQ2:} \emph{
 Do the genders of the contributors influence code review intervals for their code changes?}
\end{boxedtext}

We formalize this research question into the following two hypotheses:

\boldmath{$H2.1_0$:} \textit{There are no significant differences in review intervals between men and women.}

\boldmath{$H2.1_a$:} \textit{There are significant differences in code review intervals between men and women.}

Since identifiable gender may influence code review intervals, we pose two additional hypotheses.

\boldmath{$H2.2_0$:} \textit{There are no significant differences in code review intervals between men/women with and without gender identifiable profiles.}

\boldmath{$H2.2_a$:} \textit{There are significant differences in code review intervals between men/women with and without gender identifiable profiles.}

\vspace{4pt}
Although women did not exhibit lower productivity than men, Bosu \textit{et} al. ~\cite{bosu-esem-2019} found significantly lower participation of women in code reviews in one project. They hypothesize that, since participation in a code review is based on invitations from peers, a man may prefer other men over women contributors for his reviewing code changes~\cite{bosu-esem-2019}. Hence, our next research question is:

\begin{boxedtext}
\textbf{RQ3:} \emph{
 Do the genders of the contributors influence their participation in code reviews?}
\end{boxedtext}

We formalize this research question into the following two hypotheses:

\boldmath{$H3.1_0$:} \textit{There are no significant differences in code review participation between men and women.}

\boldmath{$H3.1_a$:} \textit{There are significant differences in code review participation between men and women.}

However, as identifiable gender may influence code review invitations, we pose the following two hypotheses.

\boldmath{$H3.2_0$:} \textit{There are no significant differences in code review participation between men/women with and without gender identifiable profiles.}

\boldmath{$H3.2_a$:} \textit{There are significant differences in code review participation between men/women with and without gender identifiable profiles.}

\section{Research method}
\label{sec:dataset}
We conducted our study on a total of 1010 projects, where 10 projects use  Gerrit~\cite{gerrit} and 1,000 projects are mined from GitHub based on a stratified sampling strategy. 
While we acknowledge the subtle difference between Gerrit-based code reviews~\cite{Bosu-Carver-ESEM:2014} and GitHub's pull request-based ones~\cite{gousios2014exploratory}, the participation and outcome aspects addressed in this study are mostly similar. For the sake of brevity, we use the term `CR' to refer to both code review and pull request hereinafter. 
The following subsections detail our approaches to selecting confounding variables relevant to our research questions, preparing datasets for  Gerrit and GitHub-based projects, computing attributes, and analyzing data.

\subsection{Variable Selection}
\label{sec:variables}
Table \ref{table:variable_descriptions} lists the three dependent, two independent, and eleven confounding variables for our study. We have grouped the confounding variables into two categories, i.e., contributor characteristics and changeset characteristics. Both of our independent variables (i.e., Gender, and Gender neutral profile) also belong to the contributor characteristics group. 
Table \ref{table:variable_descriptions}  also includes a description and rationale behind the selection for each of the variables.
We selected a confounding variable for inclusion if either a prior study has found its influence on CR outcomes~\cite{pull_based,mcintosh,lenarduzzi2021does,jeong2009improving,tao2014writing,fan2018early,barnett2015helping,thongtanunam2015should} or it was included in one of the two original studies~\cite{terrell_1,bosu-esem-2019}. 
The name of the variable inside parentheses denotes its acronym that we have used to refer to it hereinafter. 

\begin{table*}
	\caption{Descriptions and rationale of the dependent and independent variables. We group the independent variables into two categories: i) contributor characteristics and ii) changeset characteristics. For building the `CR participation model', we will use only the independent variables listed in the `contributor characteristics' group.}
	\centering \label{table:variable_descriptions}
	\resizebox{\textwidth}{!}{%
\begin{tabular}{p{2cm} p {3cm} p {6cm}}
\hline
\rowcolor[HTML]{D9D9D9} 
\textbf{Name}           & \textbf{Description}                                                                                                                                     & \textbf{Rationale}                                                                                                                                                                                                                                                                                                                                                      \\ 
\rowcolor[HTML]{ffffff} 
\multicolumn{3}{c}{\cellcolor[HTML]{ffffff}\textbf{ Independent variables}}  \\   
\multicolumn{3}{l}{\cellcolor[HTML]{C0C0C0}\textbf{Variable group: Contributor characteristics}}  \\ 
\rowcolor[HTML]{D9D9D9} 
Gender (Gender)                  & Whether a contributor is a man or woman  & This study's objective is to identify influences of gender.\\

\rowcolor[HTML]{EFEFEF} 
 Gender neutral profile (isGenderNeutral)  & If a contributor's profile is gender neutral  &   Terrell \textit{et} al. found identifiable gender influencing acceptance rates~\cite{terrell_1}. \\                                                                                                                                                                                                                                                                         
\rowcolor[HTML]{D9D9D9} 
Number of total commits (totalCommit) &  Number of code commits a   developer has made to a project &  Changes submitted by experienced developers who have higher number of commits can be  subjected to less scrutiny and may include lower number of mistakes than inexperienced developers~\cite{bosu2016process}.  Experienced developers are also more likely to be invited for code reviews~\cite{mirsaeedi2020mitigating}.\\

\rowcolor[HTML]{EFEFEF} 
Tenure (tenure)                 & Span of time developer is  contributing to the project &  Developers who are working on project for longer period time are more knowledgeable of the project's design~\cite{mirsaeedi2020mitigating}, are able to identify potential reviewers quickly than newcomers~\cite{Bosu-Carver-ESEM:2014}, and are more likely to be invited as reviewers.   
\\ 

\rowcolor[HTML]{D9D9D9} 
Review experience (revExp) & The number of pull requests that a developer has examined in the project as a reviewer              & Experienced reviewers are more likely get invited for more reviews~\cite{mirsaeedi2020mitigating}                                                           
\\

\multicolumn{3}{l}{\cellcolor[HTML]{C0C0C0}\textbf{ Variable group: Changeset characteristics}}  \\ 

\rowcolor[HTML]{EFEFEF} 
Patch size (patchSize)              & Number of lines modified/  added/ deleted in a code review request                         & Larger changes are more likely to be buggy,  require  longer time to review~\cite{Bosu-et-al-FSE:2014} and less like to be accepted~\cite{jiang2013will}. \\

\rowcolor[HTML]{D9D9D9} 
Cyclomatic complexity (cyCmplx)   & McCabe's cyclomatic  complexity                                                              & When a patch set is difficult to comprehend, it will need a longer review intervals~\cite{bacchelli2013expectations} and is more likely  to be rejected.  \\

\rowcolor[HTML]{EFEFEF} 
Number of patchset (numPatch)      & Number of revisions executed in the submitted request &  Higher number of patchsets would increase     review intervals as well as probability of rejections~\cite{Bosu-Carver-ESEM:2014}  \\

\rowcolor[HTML]{D9D9D9} 
Is bug fix (isBugFix)              & Whether a change includes a new feature or bug fix & Bug fixes may be in a urgent need and get reviewed quickly~\cite{paul2021security}.  \\

\rowcolor[HTML]{EFEFEF} 
Number of files (fileCount)         & Number of files under review in review requests    & Changes with higher number of  files involved, are more likely to be defect-prone\cite{review_2} and require longer review time. \\    

\rowcolor[HTML]{D9D9D9} 
Number of directories (dirCount)    & Number of directories in a review request where files  have been  modified     & Changes spread across a large number number of  directories  are more likely to be buggy, difficult to comprehend, and require longer review time~\cite{barnett2015helping}.  \\

\rowcolor[HTML]{EFEFEF} 
Comment volume (cmtVolume)          & Ratio of add/modified lines that are comments  &         Well commented changes are easy to comprehend~\cite{barnett2015helping} and require shorter review intervals. \\

\rowcolor[HTML]{D9D9D9} 
 Ratio of new files (ratioNew)             &  Ratio of newly created files to total number of files in a patchset     & A new file may have more issues than a file which has been under review  before~\cite{xia2015should}. As a result, a new file can have longer review interval and less  acceptance rate.  \\
\rowcolor[HTML]{C0C0C0} 
\multicolumn{3}{c}{\cellcolor[HTML]{ffffff}\textbf{Dependent Variables}}                                                                                                                                                                                                                                                                                                                                                                                                                                                                                                                                    \\
\rowcolor[HTML]{D9D9D9} 
Acceptance (isAccepted)          & Whether a code change was accepted or rejected   &  Indicates the outcome of a code review or pull request. \\

\rowcolor[HTML]{EFEFEF} 
Code review interval (reviewInteval)    & Time from beginning of the code review to the end      & Indicates whether a contributor's code gets preferential treatments in a FOSS community   \\
\rowcolor[HTML]{D9D9D9} 
Code review participation (avgReviewPerMonth)   & Average number of code review where the developer participated in the project   per month    &    Indicates whether a contributor is valued by his/her peers.                                                                                              \\ \hline
\end{tabular}

}
	\vspace{-8pt}
\end{table*}

\begin{table}
	\caption{Demographics of Gerrit projects}
	\centering \label{table:demographic-project}
	\resizebox{\linewidth}{!}{

\begin{tabular}{|l|l|r|r|}
\hline
\textbf{Project}     & \textbf{Sponsor}           & \textbf{Request mined} & \textbf{\# of developers*} \\ \hline
Android     & Google            & 637,064         & 3,351                                                                \\ \hline
Chromium OS & Google            & 1,536,071        & 5,375                                                               \\ \hline
Couchbase   & Couchbase Inc.    & 140,124        & 300                                                                \\ \hline
Go          & Google            & 47,096         & 335                                                                \\ \hline
LibreOffice & Foundation        & 127,812         & 573                                                                \\ \hline
oVirt       & Redhat Inc        & 113,465         & 292                                                                \\ \hline
Qt          & Qt Company        & 184,290        & 815                                                                \\ \hline
Typo3       & Foundation        & 65,399         & 412                                                                \\ \hline
Whamcloud   & Intel             & 21,724         & 158                                                                \\ \hline
Wikimedia   & Foundation        & 494,302         & 904                                                                \\ \hline
Total       &                   & 3,367,347      & 12,252                                                               \\ \hline
\multicolumn{4}{l}{* Number of developers, who have made at least five commits to this project.}
\end{tabular}
}
	\vspace{-8pt}
\end{table}

\subsection{Gerrit Dataset Preparation}
Due to an overlapping author, we had access to the dataset of  683,865 CRs curated in Bosu \textit{et} al.'s~\cite{bosu-esem-2019} study. Projects of this dataset were selected based on two criteria: 1) Projects are actively using Gerrit. 2) Contributors of that project reviewed at least 15,000 requests.
Since OmapZoom, one of the projects from their selection, is no longer active,  we replaced it with WikiMedia. WikiMedia also satisfies selection criteria laid out by Bosu \textit{et} al.  
Using the same Gerrit-Miner tool used in their study, we updated Bosu et al.'s dataset to include all CRs completed till 30th April 2022.
We identified the bot
accounts for exclusion using a set of keywords (e.g., `bot’, `CI’, `Jenkins’, `build’, `auto’, `devop', `hook', `workflow' and ‘travis’) followed by manual validations. Similar to Bosu \textit{et} al. , we used the Levenshtein distance between
two names to identify similar names. We performed manual reviews of the associated accounts to identify and merge the multiple accounts belonging to the same person into one account.
Similar to the original study~\cite{bosu-esem-2019}, we limited our analysis to non-casual contributors.  We also wrote another script to download and store avatar images for all the users.
Table \ref{table:demographic-project} lists the demographic of the projects. This updated dataset includes more than 3.3 million CRs, which is approximately five times more than the original study. We include a total of 12,252 non-casual developers, which is also 2.66 times more than Bosu \textit{et} al.'s.

\subsection{GitHub Dataset Preparation} 
We used the publicly available GHTorrent MySQL dump from March 2021. We imported this data dump on a local MySQL instance.
This dataset includes a total of 45.8 million users and 109.2 million pull requests. Since it is not feasible to include such a large number of PRs, we focused on selecting a  sample of 1000 projects. 
Following the recommendations of Kallamvakou \textit{et} al.~\cite{kalliamvakou2016depth},  
We defined the following selection criteria to include only software projects actively using the PR-based development model~\cite{gousios2014exploratory}. 

\begin{enumerate}
   \item A software development project using one of the popular programming languages, such as C, C++, Java, JavaScript, Python, PHP, Ruby, C\#, Scala, Lua, Go, Kotlin, Typescript, and Rust.
    \item Source code is available under an open Source license.
    \item At least 20 contributors.
    \item Actively follows pull-based development with at least 20 PRs during the last three months in our dataset.
\end{enumerate}

\vspace{4pt}
\noindent \emph{Project selection: }
To facilitate our selection, we wrote queries to compute the number of distinct contributors,  and the number of PRs for each repository. 
We hypothesized that the characteristics of gender biases may also depend on community size. To investigate this hypothesis, we divided the projects into four groups based on the number of contributors. To facilitate referencing these four groups, we named them according to community size, which is as follows:
\begin{enumerate}
   \item \emph{Small [GitHub (S)]:} Projects that have 20-100 contributors.
    \item \emph{Medium [GitHub (M)]:} Projects having 101-250 contributors.
    \item \emph{Large [GitHub (L)]:}  Projects with  251-500 contributors.
    \item \emph{Extra large [GitHub (XL)]:} Projects with more than 500 contributors.
\end{enumerate}
From each group, we randomly selected 250 projects satisfying our four selection criteria. 
Since GHTorrent dump does not include all the required attributes necessary for this study, we mined those data directly from GitHub. To ensure that all the selected projects are still available, we wrote a Python script using the PyGithub library~\cite{pygithub}. From our initial selection, we found 27 unavailable projects. We randomly selected 27 available projects to replace those unavailable ones. 

After finalizing the set of projects, we wrote queries to identify all the users who have opened at least one PR or made at least a commitment to one of the 1,000 selected projects. We found a total of 458,486 GitHub users meeting these criteria. After excluding organization and \emph{deleted} accounts based on the attributes listed in the \code{Users} table, we found a total of 310,983 users. 
GHTorrent also does not include personal information such as full name, email address, bio, blog, location,  avatar, and Twitter that may assist us in gender identification. We wrote a Python script to mine that information and populate our `user\_details table. We wrote another Python script to download the avatars of all users and store avatar images with the corresponding GitHub profile id. We used the same keyword-based filtering followed by the manual validation approach used on the Gerrit dataset, to identify and remove bot accounts from our dataset. 

Due to the extremely large size of the GHTorrent dataset, queries take a significant amount of time. Therefore, we wrote a script to create new tables with a subset of rows that are related to our 1,000 selected projects from the full tables. We put those shortened tables in a new database called `ghtorrent\_short'. 
Table \ref{table:github_projects} shows the demographics of the selected projects from our ghtorrent\_short dataset. Our dataset includes a total of 2,839,192 PRs.

Initially, we selected 250 projects from each group (GitHub(S), GitHub(M), GitHub(L), GitHub(XL)) for data analysis. Unfortunately, while downloading pull request details for each of those projects, some of the projects were missing. So, we could not take exactly 250 projects from each group, instead, it is \SI{250 \pm 2}{}. However, the total number of projects remains at 1000 for the GitHub dataset.



\begin{table}
	\caption{Overview of the selected GitHub projects}
	\centering \label{table:github_projects}
	\resizebox{\linewidth}{!}{
\begin{tabular}{|l|l|r|r|}
\hline
\textbf{Project}     & \textbf{\# of contributors}           & \textbf{PR mined} & \textbf{Total users*} \\ \hline
Github(S)    & 20-100            & 80,606         &     11,714                                                         \\ \hline
Github(M)   & 101-250            & 283,920        &  36,244                                                               \\ \hline
Github(L)   & 251-500           & 647,574        & 70,162                                                                \\ \hline
Github(XL)  & more than 500            & 1,827,092         & 192,863                                              \\ \hline
Total       &                   & 2,839,192      & 310,983                                                              \\ \hline
\multicolumn{4}{l}{* Number of developers, who have opened at least one PR to these group.}
\end{tabular}
}
	\vspace{-8pt}
\end{table}

\vspace{4pt}
\noindent \emph{Fetching pull request details:} Our study also requires several attributes of each PR that are unavailable in GHTorrent. 
We wrote a Python script using the PyGithub library~\cite{pygithub}  to mine the following attributes for each PR: commit count, the number of changed files, the sum of lines added, the sum of lines deleted, date created, date pull merged, date closed, the total number of comments, current status, merge SHA and raw URLs of the files. 
Our script also downloads all the listed files in a PR using the raw URLs and computes various attributes such as Cyclomatic complexity, number of lines, number of lines that are comments, number of methods, lines added, lines deleted, and file status.
We have added two new tables, \code{pull\_details} and \code{pull\_files} in our GHTorrent\_short schema to store that information.
Section~\ref{sec:attributes} provides more details on how each of these attributes is computed.

\subsection{Variable computation}
\label{sec:attributes}
The following subsections detail the computation procedure for the three dependent, two independent, and eleven confounding variables. 

\subsubsection{Gender (Gender)}
\label{gender-comp}
Our four-step, semi-automated gender resolution steps are adopted from the methodologies used by Terrell \textit{et} al.~\cite{terrell_1} and Bosu \textit{et} al.~\cite{bosu-esem-2019}.
Similar to those studies, we used the GenderComputer tool~\cite{gender_v} to resolve developers' genders from name and location as the first step.  In the second step, we used face detection and gender recognition machine learning models to identify genders from profile avatars. In the third step, we manually verified genders using LinkedIn and social media searches, if the first two steps failed or returned contradictory binary genders, e.g., name resolution suggests a man but avatar suggests a woman.

\begin{enumerate}[leftmargin=*]

\item \emph{Step 1: Gender resolution from name} To resolve the gender of the GitHub users, we started with using GenderComputer\footnote{https://github.com/tue-mdse/genderComputer}. This tool takes the first and last name of a person and their location if that is available. The GenderComputer tool classifies a person into one of the following four categories:  1) Male, 2) Female, 3) Unisex, and  4) Unknown. We executed this step for all the users from both GHTorrent and Gerrit datasets.

\item \emph{Step 2: Identification of gender from avatar} We used avatars downloaded from Gerrit and GitHub profiles to infer gender. If a user has not customized their avatar, an identicon is used, which is a 5×5 pixel sprite generated using the hash of the user's ID. We used  an empirically developed file size-based heuristics to identify identicons. Since we noticed that all the identicons are less than 1.5 kilobytes, we moved all the avatars less than 1.5 kilobytes to our ``identicon'' directory. However, this size-based heuristics is not entirely accurate, as a very small number of custom avatars also fall into this size group and vice versa. To fix such misclassifications, we manually inspected all the identicons. To save inspection time, we inspected the files in the thumbnail view mode of the directory. However, since we have a large number of files, loading a  directory with a large number of files in the thumbnail view is time /resource consuming. To counter this challenge, we executed a shell command to divide the ``identicon'' directory into multiple sub-directories with each directory including 1,000 files. During our thumbnail view of each sub-directory, if we noticed any file that was misclassified, we used the ``drag and drop" action to move the file to the correct directory. 

After excluding the identicons, we used the Haar Cascade human face detection algorithm from the OpenCV library~\cite{goyal2017face} on the non-identicons.
We wrote a Python script that places each non-identicon into one of the following three groups:  1) \textit{Unknown}: images with no human face, 2) \textit{single}: images with only one human face, and 3) \textit{multiple}: images with multiple human faces. After this script's execution, we noticed misclassifications ranging from 10-15\%. For correction, we used the same thumbnail-based viewing strategy used for grouping identicon /non-identicons to move misclassified images to the correct directories. 

After manual corrections, we used another Python script to automatically infer gender from the ``single" images. This script uses the OpenCV library to automatically identify rectangular sections containing faces. Next, it uses pre-trained models created by Eidinger et al. ~\cite{eidinger2014age} to classify each image as `man' or `woman'. To correct misclassifications, we repeated our thumbnail view-based manual inspection strategy. During this manual inspection phase, we also inspected the misclassifications in a full-scale view, as some images could be only determined after careful inspections. Moreover, we also noticed many avatars are photos of children. During our manual inspections, we moved avatars with children seemingly less than 13 years old to the `Neutral' directory based on an assumption that such photos are of someone other than the account holder. We choose 13 years as the threshold since GitHub policy requires a user to be at least 13 years old. During our manual inspections, we also noticed avatars using photos of famous singers, actors, cartoon characters, well-known internet memes, and athletes. We also categorized such avatars as `Neutrals', as these avatars represent users' interests rather than themselves.

Finally, we manually inspected all the avatars with multiple human faces. If an avatar includes only one person over 13 years of age, we assign that avatar to that person's gender. For example, such a photo could be a father with his son/daughter and this account belongs to the father. However, if multiple persons over 13 years were present, we classified avatars as `Neutrals'. While our gender inference for avatars were extremely time-consuming, we resorted to this process due to the unavailability of another reliable alternative. To ensure the reliability of our results, we wanted to ensure very high accuracy in this step. From GHTorrent, we identified a total of 10,081 women, 143,094 men, and the remaining with either identicons or neutral avatars.  Our replication package also includes examples from each category~\cite{replication}. 
We used the same approach to infer genders from Gerrit avatars as well. However, among ten Gerrit projects, four projects: LibreOffice, Qt, Whamcloud, and Wikimedia do not use any profile avatar. Therefore, we skipped this step for those four projects.

\item \emph{Step 3: Semi-automated resolution} For each user, we inferred their gender from two sources: first from full name and location and second from profile avatar. If both of these results match, we labeled a person's gender as that. If either of the results is `unknown' and the other one suggests man/ woman, we assign that gender to the user. If the results conflict, we check the profiles manually and conduct an additional investigation using social media (i.e., LinkedIn and Facebook) to decide. If both name and avatar-based resolutions fail for a person, we assign `Unknown' to that user. 

\item \emph{Step 4: Manual search}  
We conducted manual investigations of the unresolved profiles (i.e., `Unknown' at the end of Step 3) to determine genders. First, we exclude all the users that do not list at least one of the following information: name, email address, blog, and Twitter. For the remaining users, we used an approach similar to Bosu et al.~\cite{bosu-esem-2019}. However, since Google+ is no longer available, we replaced it with Google People API\footnote{https://developers.google.com/people}.
Moreover, since the manual resolution is extremely time-consuming, we also excluded users, who had opened less than five CRs. 
In the first step, we searched for Google people profiles using email addresses to identify photos. If a person is not found or we cannot decide based on the associated photo, we search for that person on LinkedIn using their name and email domain. If a Twitter account or a personal website is listed on a person's profile, we also inspected those. 

Although our gender resolution steps are time-consuming, we were able to achieve a high-resolution rate. For GitHub, we were able to resolve 71.5\% users. For the GitHub users, who have opened at least 5 PRs and have emails listed on their profiles, we resolved 89.6\%. For Gerrit-based projects, overall we achieved 98.8\% success for non-casual developers.
Since our study is limited to investigating gender bias in men and women, we removed user ids for whom we assigned 'Unknown' as gender. For data analysis, we coded `0' for men and `1' for women.

\end{enumerate}

Table \ref{table:gender_demographics} shows the demographics of gender for the Gerrit projects and four groups of GitHub projects.  The ratio of women varies between 4-19\%, where the highest ratio is seen in Wikimedia. Among the GitHub groups, the ratio of women increases as the project size grows. Overall, we found 7.4\% women.  

\begin{table}
	\caption{Demographics of gender in Gerrit and GitHub datasets}
	\centering \label{table:gender_demographics}
	
\resizebox{\textwidth}{!}{%

\begin{tabular}{|l|r|r|r|r|}
\hline
\textbf{Project}     & \textbf{Men}           & \textbf{Women} & \textbf{Men using} & \textbf{Women using }\\
 &       &  & \textbf{GNPs} & \textbf{GNPs}\\\hline
Android     & 2896 (86.7\%)           & 444 (13.3\%)       & 1619 (55.9\%) &    306   (68.9\%)   \\ \hline
Chromium OS &   4,466 (83.3\%)         &   897 (16.7\%)      &  3,347 (74.9\%)  &  728 (81.2\%)                        \\ \hline
Couchbase   &  265 (88.3\%)    &    35 (11.7\%)    & 160 (60.4\%)  &       28 (80.0\%)                                                   \\ \hline
Go          &    313 (93.4\%) &  22 (5.6\%)    &  188 (60.1\%)      &  18 (81.8\%)                                                             \\ \hline
LibreOffice$\ddagger$ &   518 (90.9\%)     &   52 (9.1\%)       &  -  &     -                                                        \\ \hline
oVirt       &   258 (88.4\%)     &    34 (11.6\%)    & 208 (80.6\%)    &   29 (85.3\%)                                                 \\ \hline
Qt$\ddagger$          &  777 (95.3\%)      &  38 (4.7\%)       &   -   &     -                                                     \\ \hline
Typo3       &  392 (95.2\%)      &   20 (4.8\%)       &   195 (49.7\%)   &   13 (65.0\%)                                           \\ \hline
Whamcloud$\ddagger$   &  145 (91.8\%)         &  13 (8.2\%)      &  -       &  -                                                      \\ \hline
Wikimedia$\ddagger$   &  637 (81.3\%)      &   147 (18.7\%)       &  -       &   -                                                     \\ \hline
Github(S)   &    9,064 (94.1\%)         & 566 (5.9\%)        & 2,929 (32.3\%)   &   277 (48.9\%)            \\\hline
Github(M)   &    26,279 (93.3\%)          & 1,886 (6.7\%)       & 8,810 (33.5\%)   &   934 (49.5\%)           \\ \hline
Github(L)   &    49,561 (93.2\%)         & 3,604 (6.8\%)       & 17,002 (34.3\%)  &   1,853 (51.4\%)          \\ \hline
Github(XL)   &    121,765 (92.7\%)          & 9,639 (8.5\%)      & 42,241 (34.7\%)  &   4,887 (50.7\%)          \\ \hline \hline

\textbf{Total: } & \textbf{217,336 (92.6\%)} &	\textbf{17,397 (7.4\%)} &	\textbf{76,699 (35.3\%)} &	\textbf{9,073 (52.2\%)} \\ \hline

\multicolumn{5}{p{11cm}}{$\ddagger$ - since these projects do not provide an option to use avatar, we consider all profiles as gender neutrals.}
\end{tabular}
}
	\vspace{-8pt}
\end{table}

\subsubsection{Gender neutral profile (isGenderNeutral)}
\label{profile-comp}
Prior study shows that women often use gender-neutral profiles (GNP) to avoid unwanted attention. Terrell \etal also showed that women have higher acceptance rates when their gender is not identifiable. To test our formulated hypotheses: $H1.2$, $H2.2$, $H3.2$, we need to identify if a profile is gender neutral. If users' genders are identifiable from their avatars, we consider them as gender identifiable profiles (GIP) and put down 0’s as the values for \textit{isGenderNeutral} variable. Otherwise, we considered their profiles as gender-neutral profiles (GNP) and put 1’s for this variable.

We like to mention that our GNP /GIP  resolution strategy is different from Terrell \textit{et} al.'s.  To identify if a profile uses identicon, Terrell \textit{et} al. checked for image dimensions using ImageMagick. If a profile used an identicon, and the gender inference tool classified its user name as `Unknown', they used a panel of three people to look into the login and user name for 10 seconds to determine whether they can reliably guess the gender. However, they added this manual step only for 3,000 randomly selected profiles. During our study, we noticed two shortcomings of this strategy. First, we noticed many custom avatars with the same dimension as identicons. Therefore, a non-negligible number of identicons identified based on this strategy are false positives. Second, custom avatars can also be gender-neutral (e.g., pets, celebrities, kids, inanimate objects). Third, inferring gender from a name also requires familiarity with the origin. For example, most people from the Indian subcontinent may easily infer the name `Sayma' belongs to a woman. However, people from different origins may have difficulties. Moreover, even though a name-to-gender inference tool can reliably guess gender from a name, CR participants are less likely to use such tools during their collaborations.
Therefore, to avoid such limitations, we considered only avatars as sources to identify gendered profiles, i.e., we marked a profile as GIP only if we can reliably identify its owner's gender using the associated avatar. 
This step is similar for both Gerrit and GitHub datasets. Since four of the Gerrit projects (i.e., LibreOffice, Qt, Whamcloud, and WikiMedia)  do not use profile avatars, values for this variable are  0’s for all of the users from those projects. Table \ref{table:gender_demographics} also shows the distributions of developers from various projects using GNPs.  We also noticed the majority of women (52.2\%) prefer GNPs, while almost two third men prefer GIPs. 

\subsubsection{Commit experience (totalCommit)} For each CR /PR, we count code commit experience as the number of prior code commits for the current project. We wrote a SQL script to compute this attribute.

\subsubsection{Author tenure months (tenure)}
\label{tenure}
We measured the tenure of a developer from the difference between the timestamp of the first submitted CR  and the most recently submitted ones for a particular project. We first computed the differences in terms of the number of days. After dividing by 30, we take the ceiling values to convert those differences into the number of months. 

\subsubsection{ Review experience (revExp)} 
For each CR /PR, we measure the review experience of a participant by computing the number of distinct CRs that he/she has participated in, where he/she was not the owner (or creator of PR on GitHub).
We wrote a SQL script to compute this attribute.

\subsubsection{Patch size (patchSize)}
We first mined code churn for individual files (i.e., file\_code\_churn under a CR. We computed patch size using the following formula: 

    $patchSize =  \sum_{i=1}^{n}file\_code\_churn_{i}$, where  a CR  includes n files and $file\_code\_churn_i$ denotes the code churn of $i_{th}$ file.

\subsubsection{Cyclomatic complexity (cyCmplx) }
\label{cplx}
We wrote a python script that downloads all files included in a CR  and uses the {\tt Lizard} python library~\cite{lizard} to compute cyclomatic complexity for each file. For the regression model, we compute the average cyclomatic complexities of all files included in a CR.

\subsubsection{Number of patchset (numPatch)} For Gerrit, this variable denotes the number of review iterations made before a decision was made for a CR. For GitHub, this variable denotes the number of commits included in a PR.

\subsubsection{Is bug fix (isBugFix)}
\label{bug-fix}
We used a heuristic-based approach similar to the one proposed by Zafar \textit{et} al. to identify whether a CR is for a bug fix~\cite{zafar2019towards}. To identify CRs with bug fixes, we checked the description of the CR for links to bug repositories. Every Gerrit-based project uses a different style to link bug reports. We empirically investigated each Gerrit project to create regular expressions to identify its bug links. For example, CRs associated with bug fixes for oVirt include `Bug-Url: https://bugzilla.redhat.com/([0-9]+)' in descriptions.
For GitHub-based projects, we first marked bug-associated labels using a regular expression search. Such labels typically include keywords such as  `bug' and `defect'. We
checked if a PR is linked with an issue and that issue is assigned one of the bug-related labels. 

\subsubsection{Number of files (fileCount)}
We wrote a Python script using the Pygithub library~\cite{pygithub} to mine the number of files included in a PR. Our Gerrit miner also mines this information for all CRs.

\subsubsection{Number of directories (dirCount)}
We wrote an SQL query to extract directory names from file paths. Another query was used to compute the number of distinct directories in a CR.

\subsubsection{Comment volume (cmtVolume)}
We wrote a Python script that downloads exact versions of all the files included in a CR and computes the number of lines (nLOC) and the number of lines that are comments (nCOM) using the Lizard library~\cite{lizard}. We compute the average comment volume based on the following formula.

$cmtVolume  = \frac{1}{n} *\sum_{i=1}^{n} \frac{nCOM_i} {nLOC_i}$ 

where  a CR  includes n files and $nCOM_i$ and $nLOC_i$ denote nCOM and nLOC of the $i_{th}$ file respectively.

\subsubsection{Ratio of new files (ratioNew)}
Gerrit miner already mined whether a file was new or modified. Our Python script mined the same for all the PRs. We wrote a SQL script to compute the total number of files (i.e., numNewFile) that are new in a review. We compute this attribute as:
$ratioNew = \frac{numNewFile}{fileCount}$

\subsubsection{Acceptance (isAccepted)}
\label{acceptance}
Gerrit miner downloads the status of CRs as one of the following: `New', `Merged', or `Abandoned'. We excluded all the CRs with the status `New', since those are ongoing. For the reviews with status marked as `Merged' and `Abandoned, we set \code{isAccepted} =1 and \code{isAccepted} =0 respectively.

For each PR, we mined timestamps for `merged\_at' and `closed\_at'. We excluded open PRs (i.e. `closed\_at' = \code{NULL}). We noticed that if a PR is closed without merging, its `merged\_at' is set as \code{NULL}. Therefore, for the PRs with `merged\_at' =\code{NULL}, we set \code{isAccepted} =0. For the remaining PRs, we set \code{isAccepted} =1.

\subsubsection{Review interval (reviewInterval)}

For each PR, we computed the number of seconds between the time when the PR was created (i.e. \code{created\_at}) and the time when that PR was closed (i.e., \code{closed\_at}).  Similarly, for Gerrit-based reviews, we used differences between created and last updated. By dividing the review interval by 3600, we computed the review interval in terms of hours.

\subsubsection{Code review participation (avgReviewPerMonth):}
For each reviewer, we computed the average number of CRs over their tenure with a project. This variable is the dependent variable for RQ3. We take monthly snapshots of this variable and five independent variables (i.e., participant characteristics) over the tenure of a participant. Such snapshots are tuples listing \code{(tenure, gender, isGenderNeutral, totalCommit, \\ reviewExp, avgReviewPerMonth)}.
We took monthly snapshots instead of one single average number since the former allows us to investigate whether women had different CR participation trajectories during their tenure than men. For example, women may start slow in terms of review participation but once recognized may catch up. 

To illustrate our computation, let us assume that Alice has been contributing to the Go project for 4 months. She is a woman with a gender-neutral profile. During the months of her tenure, she committed 5, 3, 7, and 9 changes respectively.  Her number of CR participation during these months is 2, 4, 3, and 7 respectively.  Therefore, her average CRs are 2, 3, 3, and 4 respectively.  To build our regression model, we include four data points for Alice, which are : 
( 1, Woman, Yes, 5, 2, 2), ( 2, Woman, Yes, 8, 9, 3), ( 3, Woman, Yes, 15, 9, 3), ( 4, Woman, Yes, 24, 16, 4).

\subsection{Regression model training and analysis}
\label{sec:regression_model}

We use \textit{multivariate regression modeling} techniques to evaluate our hypotheses. Linear regression and logistic regression are commonly used to inspect relations among one dependent variable and one or more independent variables when the dependent variable is scalar and binary respectively~\cite{Regression}. To develop \textit{Code acceptance} models, we use logistic regression, since the dependent variable \textit{isAccepted} is dichotomous.  To analyze \textit{Code Review Interval} and \textit{Code Review Participation}, we use linear regressions, since the dependent variables  (i.e., \textit{reviewInterval} and \textit{avgReviewPerMonth}) are scalars.  We have grouped the independent variables into two categories: i) contributor characteristics and ii) changeset characteristics. While both groups of independent variables are used to train \textit{Code acceptance} and \textit{Code Review Interval} models, we use only the  `contributor characteristics' group to train \textit{Code Review Participation} models. 
Since changeset characteristics are specific to a particular code change, those may not be influential in deciding how many reviews a person participates in during a period. Table~\ref{table:variable_descriptions} describes the variables for our models.

Recent SE studies~\cite{mcintosh,bosu_2} inspired us to adopt the models' construction and analysis approach proposed by Harrell Jr. to validate our proposed hypotheses~\cite{harrell2015regression}. Harrell's approach enables us to formulate non-linear relationships among variables while being aware of the possibility of overfitting (i.e., where the model performs very well for the training dataset but poorly otherwise)\cite{harrell2015regression}. 
We detail our five-step model development and analysis approach in the following. 
 
\subsubsection* {Step 1: Normality adjustment} 
We found skewed distributions for three independent variables (i.e., \code{totalCommit}, \code{revExp}, and \code{patchSize}) and one dependent variable (i.e., \code{reviewInterval}. 
Following recommendations from prior CR studies\cite{mcintosh,Thongtanunam} to include such variables, we apply log transformations(i.e.,  $log_{e}(x)$).
   
\subsubsection* {Step 2: Correlation and redundancy analysis} 
\begin{figure}
    \includegraphics[width=1.\textwidth]{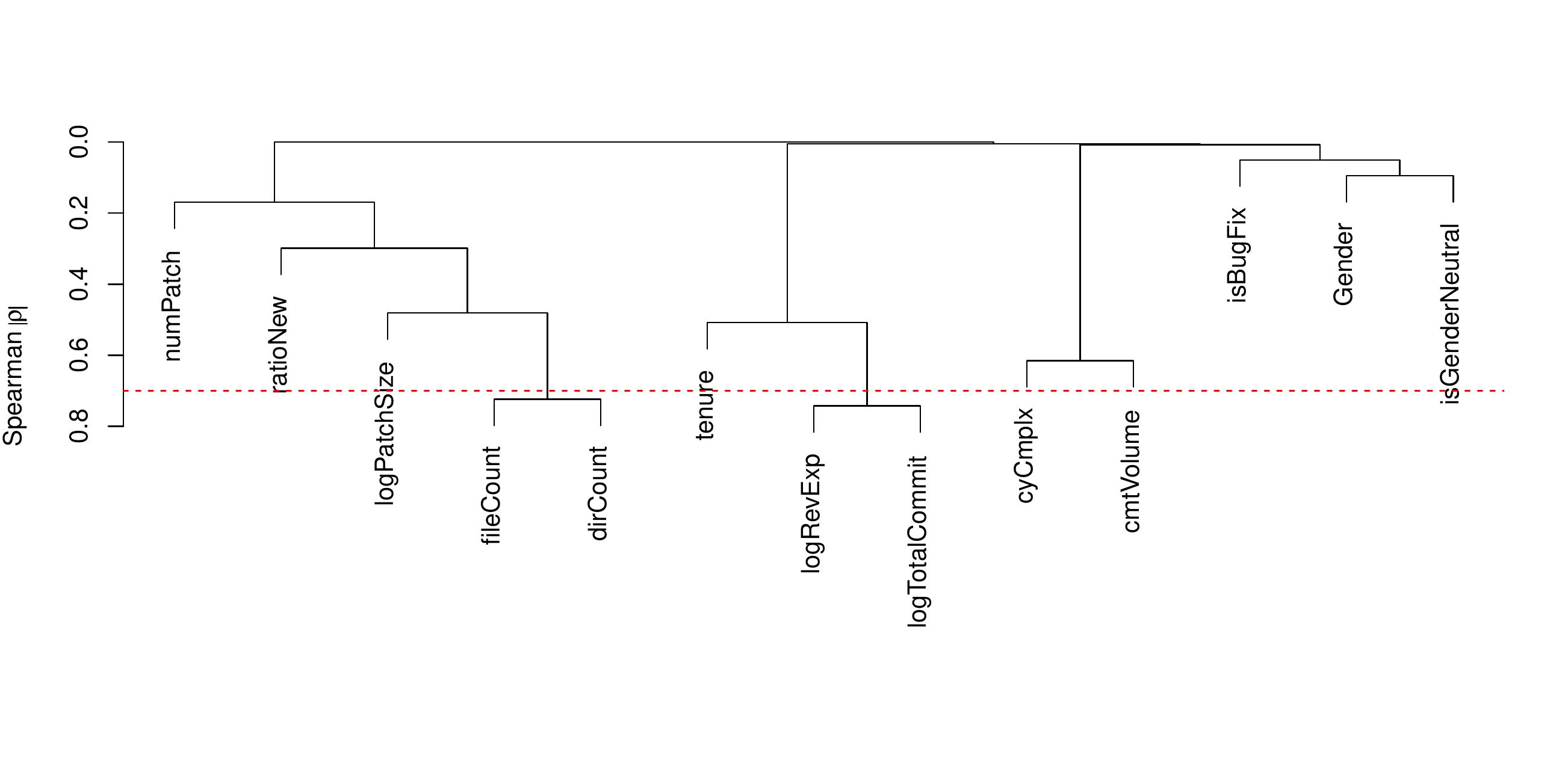}
\caption{Variable clustering for the Go project for training code acceptance model}
\label{go-variable-cluster}
\end{figure}

Before training regression models, we must assess multicollinearity, which occurs when two or more independent variables are highly correlated with one another in a regression model. Without such assessment, a model may not be able to reliably determine the effects of individual independent variables~\cite{mansfield1982detecting}.
We conduct correlation analysis among the independent variables of the models to identify highly correlated variables.  We execute the Spearman rank correlation tests between all pairs of independent variables (\(\rho \)) to construct hierarchical overviews of the correlated variables. Among the variables residing in the same sub-hierarchy and having \(\mid\)\(\rho\) \(\mid\) $<$ 0.7, we select only one for the final regression model. Prior studies in software engineering also considered 0.7 as the threshold value for identifying redundant variables \cite{mcintosh} \cite{Thongtanunam} \cite{bosu_2}.
We repeated this step for each of the 14*6 = 84 regression models that we trained.  Figure~\ref{go-variable-cluster} shows our variable clustering for our acceptance model from the Go dataset. The red trend line indicates our cutoff correlation (i.e., 0.7). If the cluster meeting point for two or more variables is below this trend line, we select only one variable (i.e., the one with the highest correlation with the dependent variable) from that cluster. In this example, \code{fileCount} and \code{dirCount} are highly correlated. Among these two, we took only \code{dirCount} for training this project’s regression model, since it has a higher correlation with the dependent variable \code{isAccepted}. Similarly, from the correlated pair (\code{logRevExp}, \code{logTotalCommit} ), we took only \code{logRevExp}. With the surviving variables, we computed the variance inflation factor (VIF). None of the predictors have VIF greater than 2.5, suggesting that multicollinearity was not a substantial threat.

 \subsubsection* {Step 3: Modeling non-linear relationships}
 Restricted cubic splines (RCS) are often used during regression modeling to model non-linear relationships between explanatory variables and outcome~\cite{durrleman1989flexible}. In this transformation strategy, the range of values of an independent variable is split up, with ``knots” defining the end of one segment and the start of the next. During regression,  if more degrees of freedom or independent variables are used than the dataset can afford, regression models may overfit. An overfitted model fails to show actual relationships among dependent and independent variables. We calculate the budget for total degrees of freedom for our code review dataset, following Harrell Jr. 's recommendations of at most n/15 total degrees of freedom, where n is the number of rows in a dataset~\cite{harrell_2,harrell_3}. For example, our smallest dataset (i.e., Whamcloud) includes 21,555 rows after excluding unresolved participants. Therefore, we can allocate at most 1,437 degrees of freedom without overfitting.  The maximum number of degrees of freedom allocated for model training was 27. Therefore, our models do not have overfitting threats.

However, we cannot allocate additional degrees of freedom to dichotomous variables such as \code{Gender}, \code{isGenderNeutral}, and \code{isBugFix}. For the remaining independent variables that survived multicollinearity analysis, we allocated four degrees of freedom using the \code{rcs} method from the \code{rms} R package.

\subsubsection* {Step 4: Training regression models} Equations (1), (2), and (3) denote regression models for \textit{Code acceptance}, \textit{Code review Interval}, and \textit{Code review participation} respectively.
To test the interaction between \code{Gender} and \code{isGenderNeutral} (i.e., $H1.2$, $H2.2$, and $H3.2$), we replaced  \code{Gender} and \code{isGenderNeutral} factors with \code{Gender * isGenderNeutral},  a four-level interaction factor in each of the following three equations.
We use the \code{lm} method from the R \code{stats} package to fit the models. Since the\textit{Code acceptance} model has binary outcomes, we use \code{binomial} linking function. For the remaining two models, we use \code{gaussian} aka `identity' linking. 
However, these equations differed among the projects, if some variables were found to be redundant during Step 2.

 \vspace{-6pt}
 \begin{equation}
  \begin{aligned}[b]
   &  \tt isAccepted \thicksim rcs(log(totalCommit),4) + rcs(log(patchSize),4) \\
   & \tt + rcs(log(revExp),4)  + rcs(tenure,4) +  rcs(cyCmplx,4) 
   \\ & \tt + rcs(numPatch,4) + isBugFix + rcs(dirCount,4) + rcs(cmtVolume,4) \\
  & \tt +  rcs(ratioNew,4) + rcs(fileCount,4)  + Gender +  isGenderNeutral\\
  \end{aligned}
\end{equation}

  \vspace{-6pt}
 \begin{equation}
  \begin{aligned}[b]
  & \tt log(reviewInterval) \thicksim rcs(log(totalCommit),4) + rcs(log(patchSize),4) \\
   & \tt + rcs(log(revExp),4)  + rcs(tenure,4) +  rcs(cyCmplx,4) 
   \\ & \tt + rcs(numPatch,4) + isBugFix + rcs(dirCount,4) + rcs(cmtVolume,4) \\
  & \tt +  rcs(ratioNew,4) + rcs(fileCount,4)  + Gender +  isGenderNeutral \\
  \end{aligned}
\end{equation} 

\vspace{-6pt} 
 \begin{equation}
  \begin{aligned}[b]
  & \tt avgReviewPerMonth \thicksim rcs(tenure,4) + rcs(log(totalCommit),4) + Gender \\
 & \tt+ rcs(log(revExp),4) +isGenderNeutral 
  \end{aligned}
\end{equation}


\subsubsection* {Step 4: Assessment of model performance} There are several metrics to assess the goodness of fit for regression models. 
In this study, we use Veall and Zimmerman Pseudo  $R^2$~\cite{veall1994evaluating} for the logistic regression models, since prior research~\cite{smith2013comparison} found this measure having closer correspondence to ordinary least square $R^2$. Veall and Zimmerman's index is defined as follows:

\[ R^2_{vz} = 
\frac{2 [LL(Null) - LL(Full)]}{ 2 [LL(Null) -LL(Full)]+ N} . \frac{2LL(Null) -N}{ 2LL(Full)}\].
Where $LL(full)$ is the log-likelihood of the full model and $LL(null)$ represents the same for a model eliminating all the independent variables (i.e., intercept-only model).

For linear regression models, we measure Adjusted $R^2$~\cite{yin2001estimating} to assess goodness of fit.

\subsubsection* {Step 5: Interpretation of results:}

For the logistic regression model (i.e., H1), we compute the Odds Ratio (OR) to estimate the impact of our variables of interest (i.e., \code{Gender}, and \code{isGenderNeutral}).  OR represents the odds of an outcome will occur with the presence of a factor, compared to the odds of the same outcome occurring in its absence \cite{fulton2012confusion}. OR is very useful in estimating the impact if a variable of interest is dichotomous.   OR = 1 indicates no impact of an independent variable on the dependent variable. A value of OR $>$ 1 indicates that, with the increase of that particular independent variable, the dependent variable has higher odds to occur, and OR $<$ 1 indicates the opposite.

For the linear regression models (i.e., H2 and H3), interpretation is more straightforward. For these models, the regression coefficient of an independent variable indicates the number of changes to the dependent variable with its 1 unit of change.  A positive coefficient indicates positive impacts on the dependent variable and vice versa.  Moreover, a larger absolute value for an independent variable's coefficient indicates a larger impact.

\section{Results}
\label{sec:result}

We present the results of our study in this section for the proposed research questions. For each question, we address analysis results and their interpretation in terms of evaluation metrics. As our study is limited to men and women, bias can work in two directions. If men are favored, we colored that attribute with blue and if women are favored, we colored the cell with pink. 

    \subsection{Gender vs. Code acceptance (RQ1) }
 \begin{boxedtext}
\revised{\textbf{RQ1:} \emph{
 Do the genders of the contributors influence acceptance of their code changes?}}
\end{boxedtext}

\begin{table*}
	\caption{Acceptance rates for men, women with gender identifiable profiles (GIP) and gender-neutral profiles (GNP)}
	\centering \label{table:avg_accep}
	
\begin{tabular}{|l|r|r|r|r|r|r|}
\hline
\textbf{Project} &  \multicolumn{2}{c|}{\textbf{Overall}} & \multicolumn{2}{c|}{\textbf{ GIP}} & \multicolumn{2}{c|}{ \textbf{GNP}} \\ \hhline{~------}
&  \textbf{Men} & \textbf{Women} & \textbf{Men} & \textbf{Women} & \textbf{Men} & \textbf{Women} \\ \hline
 Android & 86.48 & 81.26 & 84.80 & 86.67 & 87.98 & 77.98   \\ \hline 
 Chromium OS & 86.34 & 87.40 & 87.32 & 85.26 & 86.06 & 87.79   \\ \hline 
 Couchbase & 87.15 & 86.35 & 88.02 & 79.28 & 86.31 & 86.82   \\ \hline 
 Go & 88.36 & 85.94 & 91.40 & 75.02 & 85.85 & 91.12   \\ \hline 
 Libreoffice & 92.67 & 83.55 & $-$ & $-$ & $-$ & $-$   \\ \hline 
 oVirt & 86.85 & 85.23 & 87.28 & 88.63 & 86.69 & 83.65   \\ \hline 
 Qt & 83.95 & 87.77 & $-$ & $-$ & $-$ & $-$   \\ \hline 
 Typo3 & 91.33 & 90.20 & 91.87 & 91.40 & 90.02 & 80.12   \\ \hline 
 Whamcloud & 75.12 & 75.86 & $-$ & $-$ & $-$ & $-$   \\ \hline 
 WikiMedia & 91.07 & 88.52 & $-$ & $-$ & $-$ & $-$   \\ \hline 
 Github(S) & 86.34 & 86.14 & 86.78 & 86.86 & 85.36 & 85.55   \\ \hline 
 Github(M) & 78.46 & 82.40 & 77.92 & 83.35 & 79.57 & 81.21   \\ \hline 
 Github(L) & 80.04 & 82.81 & 79.54 & 84.79 & 81.02 & 79.76   \\ \hline 
 Github(XL) & 69.98 & 73.65 & 71.20 & 76.52 & 67.76 & 71.19   \\ \hline 
\end{tabular}

	\vspace{-8pt}
\end{table*}

\begin{table*}
	\caption{ Do gender/gender-neutral profiles of the contributors influence acceptance of their code changes?}
	\centering \label{table:RQ1}
	\resizebox{\linewidth}{!}{%
\begin{tabular}{| l | l | l | l || l | l |l |} 
\hhline{~~-----}
\multicolumn{2}{c}{}  &\multicolumn{5}{|c|}{\textbf{Values indicate \{Odds ratio\}$^{p-value}$ }}\\ \hline
\textbf{Project} & \textbf{Pseudo } & \multicolumn{1}{c|}{\textbf{Gender}} & \multicolumn{1}{c||}{\textbf{Neutral}}  & \multicolumn{1}{c|}{\textbf{Women}} & \multicolumn{1}{c|}{\textbf{Men}} & \multicolumn{1}{c|}{\textbf{Women}} \\
 &  $R^2$ &  & \textbf{profile}   &   \textbf{with GIP}   &  \textbf{ with GNP}   & \textbf{with GNP}   
     \\ [0.5ex] 
 \hline
 Android & 0.172 & \biasToWomen{1.08}$^{***}$ & 0.85$^{***}$ & 1.42$^{***}$ & 0.87$^{***}$ & 0.66$^{***}$  \\ \hline 
Chromium OS & 0.419 & \biasToMen{0.96}$^{***}$ & 0.96$^{***}$ & 0.84$^{***}$ & 0.95$^{***}$ & \biasToGendered{1.17}$^{***}$  \\ \hline 
Couchbase & 0.319 & \biasToMen{0.73}$^{***}$ & 0.84$^{***}$ & 0.53$^{***}$ & 0.83$^{***}$ & \biasToGendered{1.41}$^{*}$  \\ \hline 
Go & 0.552 & \biasToMen{0.62}$^{***}$ & {1.09}$^{*}$ & 0.17$^{***}$ & 0.75$^{***}$ & \biasToGendered{8.73}$^{***}$  \\ \hline 
 LibreOffice & 0.459 & \biasToMen{0.51}  $^{***}$ &  $-$ &  $-$ &  $-$  & $-$   \\ \hline 
oVirt & 0.133 & \biasToMen{0.81}$^{***}$ & {1.03} & 1.02 & 1.05$^{*}$ & 0.72$^{***}$  \\ \hline 
 Qt & 0.060 & \biasToWomen{1.55}  $^{***}$ &  $-$ &  $-$ &  $-$  & $-$   \\ \hline 
Typo3 & 0.269 & {0.95} & 0.99 & 0.99 & 1 & 0.83  \\ \hline 
 Whamcloud & 0.380 & \biasToMen{0.84}  $^{*}$ &  $-$ &  $-$ &  $-$  & $-$   \\ \hline 
 WikiMedia & 0.057 & \biasToMen{0.92}  $^{***}$ &  $-$ &  $-$ &  $-$  & $-$   \\ \hline 
Github (S) & 0.107 & \biasToWomen{1.15}$^{**}$ & 0.88$^{***}$ & 1.23$^{**}$ & 0.89$^{***}$ & 0.87  \\ \hline 
Github (M) & 0.077 & \biasToWomen{1.24}$^{***}$ & {1.12}$^{***}$ & 1.38$^{***}$ & 1.15$^{***}$ & 0.79$^{***}$  \\ \hline 
Github (L) & 0.087 & \biasToWomen{1.22}$^{***}$ & {1.06}$^{***}$ & 1.41$^{***}$ & 1.09$^{***}$ & 0.71$^{***}$  \\ \hline 
Github (XL) & 0.052 & \biasToWomen{1.27}$^{***}$ & 0.83$^{***}$ & 1.34$^{***}$ & 0.84$^{***}$ & 0.9$^{***}$  \\ \hline

\multicolumn{7}{p{11cm}}{Cells in \bluebg{blue} backgrounds represent significant biases favoring men and cells in \pinkbg{pink} backgrounds represent biases favoring women.  *** , **, and *  represent statistical significance at $p <$ 0.001, $p <$ 0.01, and $p <$ 0.05 respectively. No values among columns indicate statistically insignificance. For the gender neutral hypotheses test (i.e., H1.2), \graybg{gray} background suggests  a significant higher odds of women with neutral profiles getting  their code accepted.  }
\end{tabular}
}

	\vspace{-8pt}
\end{table*}

Table \ref{table:avg_accep} shows the average acceptance rates for groups formed based on gender, GIP, and GNP. Acceptance rates between men and women differ by 1-9\% among the projects. We noticed the largest difference in the LibreOffice project. 
Similar to Terrell \textit{et} al.~\cite{terrell_1}, we noticed between 2-5\%  higher acceptance rates for women across three of the four GitHub project groups.
For the GNP groups,  11 out of the 14 cases indicate women with GNPs having lower acceptance rates than women with GIPs; while three projects (i.e., Chromium OS, Couchbase, and Go) indicate the opposite. However, in contrast to Terrell \textit{et} al., we did not find women with GNPs having higher acceptance rates on GitHub. However, this difference may be due to our different heuristics to determine whether a profile is a GNP. 

Table~\ref{table:RQ1} shows the results of our code acceptance logistic regression models. The `Pseudo $R^2$' column in Table \ref{table:RQ1} shows the performances of the models estimated using Veall and Zimmermann index~\cite{veall1994evaluating}.
Columns 3-7 in Table~\ref{table:RQ1} indicate the measure of association between our factors of interest and code acceptance based on OR. We also mark the statistical significance of those associations using three markers, where  *** $\implies p <$ 0.001, ** $\implies p <$ 0.01, and * $\implies p <$ 0.05.  
 Since we code men with 0's and women with 1's, an  OR $>$ 1 (i.e., Gender column) implies that the probability of a CR's acceptance increases when gender changes from 0 to 1 (i.e., Man to woman) while other factors remaining unchanged and vice versa.
Similarly, an OR $>$ 1 for \code{Neutral profile} indicates higher odds of a CR's acceptance with GNPs.
To investigate interactions between the two key independent factors (i.e., gender and GNP), we trained a second regression model for each project, where we replace \code{Gender} and \code{isGenderNeutral} with \code{Gender} * \code{isGenderNeutral}. This interaction factor has four levels indicating i) men with GIPs, ii) women with GIPs, iii) men with GNPs, and iv) women with GNPs. The second regression model for a project has almost identical goodness of fit as the first one, as we are only replacing two dichotomous variables with a four-level product variable while preserving other factors.  These interaction models help us to identify if GNPs /GIPs have different associations with code acceptance for men and women. For these interaction models, we consider men with GIPs as the reference group. Therefore, the OR values reported in columns 5-7 of Table~\ref{table:RQ1} indicate odds of code acceptance with respect to men with GIPs.  For example, for the Android project, the odds of code acceptance for women with  GIPs and women with GNPs are 1.42 and 0.66 respectively with respect to men with GIPs, if other factors remain unchanged.

\subsubsection{Results of H1.1}
\noindent\textit{Gerrit dataset:} Results from Table \ref{table:RQ1} indicate that significant gender biases exist among nine out of the ten projects, with  Typo3 being the only exception. Hence, nine projects support $H1.1_a$, and the null hypothesis ($H1.1_0$) cannot be rejected for the remaining one. 
Based on \code{OR} values, seven out of those nine projects favor men. Three (i.e., Couchbase, Go, and LibreOffice) out of those seven projects indicate that women have less than three-fourths odds of getting code accepted with respect to men, even if all other factors remain unchanged. 
On the other hand, two projects (i.e., Android and Qt) show significantly higher odds favoring women. 

\vspace{4pt}
\noindent \textit{GitHub dataset:} 
Our results also suggest significant gender biases among all four GitHub groups (i.e., support for  $H1.1_a$). The OR values suggest significantly higher odds of acceptance for PRs submitted by women on GitHub. 

\subsubsection{Results of H1.2}
\noindent\ \textit{Gerrit dataset:}
As we have already indicated, we computed the\code {isGenderNeutral} attribute only for six projects, since the remaining four projects do not support avatars. OR values under the `Neutral profile' column suggest that developers with GNPs have significantly lower odds of getting code accepted in three out of those six projects, while only one project (i.e., Go) indicates the opposite.  
OR values under \code{Woman with GIP} indicate that women from this group have significantly lower odds of acceptance than men with GIPs in three projects (i.e., Chromium OS, Couchbase, and Go), while Android shows the opposite correlation. 

All six projects, except Typo3, indicate significant differences in acceptance rates between men with GIPs and men with GNPs (i.e., supporting $H1.2_a$). 
Men with GNPs have decreased odds of code acceptance in Android, Chromium OS, Couchbase, and Typo3, but have increased odds in oVirt.
 
Our comparisons between women with GIPs and women with GNPs indicate significant differences  (i.e., support for $H12._a$) in five projects.  In three projects (i.e., Chromium OS, Couchbase, and Go), women with GNPs had significantly higher odds of acceptance with respect to women with GIPs. Women with GNPs also have significantly higher acceptance rates than women with GIPs among these three projects (Table~\ref{table:avg_accep}. Therefore, GNPs increase the odds of acceptance for women in three projects. Surprisingly, the odds are almost eight times higher in Go. Our further investigation found, most of the experienced women from Go prefer using  GNPs, while less experienced women use GIPs. This particular difference between experienced and less-experienced women in avatar selection results in surprisingly higher odds of acceptance for women with GNPs.   While we do not speculate on the underlying reason for selecting GNPs, interviews of women from Go may shed more light.
On the other hand, we noticed the opposite association in two projects (i.e., Android, and oVirt), as women with GNPs have lower odds of acceptance than both men with GIPs and women with GIPs. 
In summary, five\footnote{ For our second sub-hypotheses (i.e, $H1.2_a$, $H2.2_a$, and $H3.2_a$ ), we consider a project as supporting if either of the two gender groups indicates differences between members with and without GIPs.} out of the projects show significant differences in acceptance rates between men/women with and without gender-identifiable profiles and therefore support $H1.2_a$.

\vspace{4pt}
\noindent \textit{GitHub dataset:} 
We also noticed significant associations between GNP and code acceptance among all four GitHub groups.
However,  OR values under `Neutral profile' suggest that among the projects belonging to  S and  XL  groups, participants with GNPs have lower odds of acceptance than participants with GIPs. On the contrary, projects belonging to  M and L groups show the opposite. 
Results from our interaction regression models suggest that women with GIPs have significantly higher odds of acceptance than men with GIPs across all four GitHub groups. 
Women with GNPs have lower odds than men with GIPs in projects belonging to three of the GitHub groups (i.e.,  M, L, and XL). We also noticed significant differences between men/women with GIPs and GNPs across all four groups and therefore supports for $H1.2_a$. For women, GNPs significantly decreased the odds of code acceptance across all four groups. However, for men, GNP  is associated with significantly increased odds in two groups (i.e., M, and L) but has opposite odds in the two remaining groups.

 \begin{boxedtext}
 \emph{We accept $H1.1_a$ for nine out of the ten Gerrit projects and fail to reject the null hypothesis $H1.1_0$ for the remaining one (i.e., Typo3). We tested $H1.2$ for six Gerrit projects. Among those $H1.2_a$ is supported for five projects and not supported for the remaining one (i.e., Typo3). We also found support for both  $H1.1_a$ and $H1.2_a$ across all four GitHub project groups. }

\end{boxedtext}

\subsection{Gender vs. code review interval (RQ2)}

\begin{figure}
	\includegraphics[width=\linewidth]{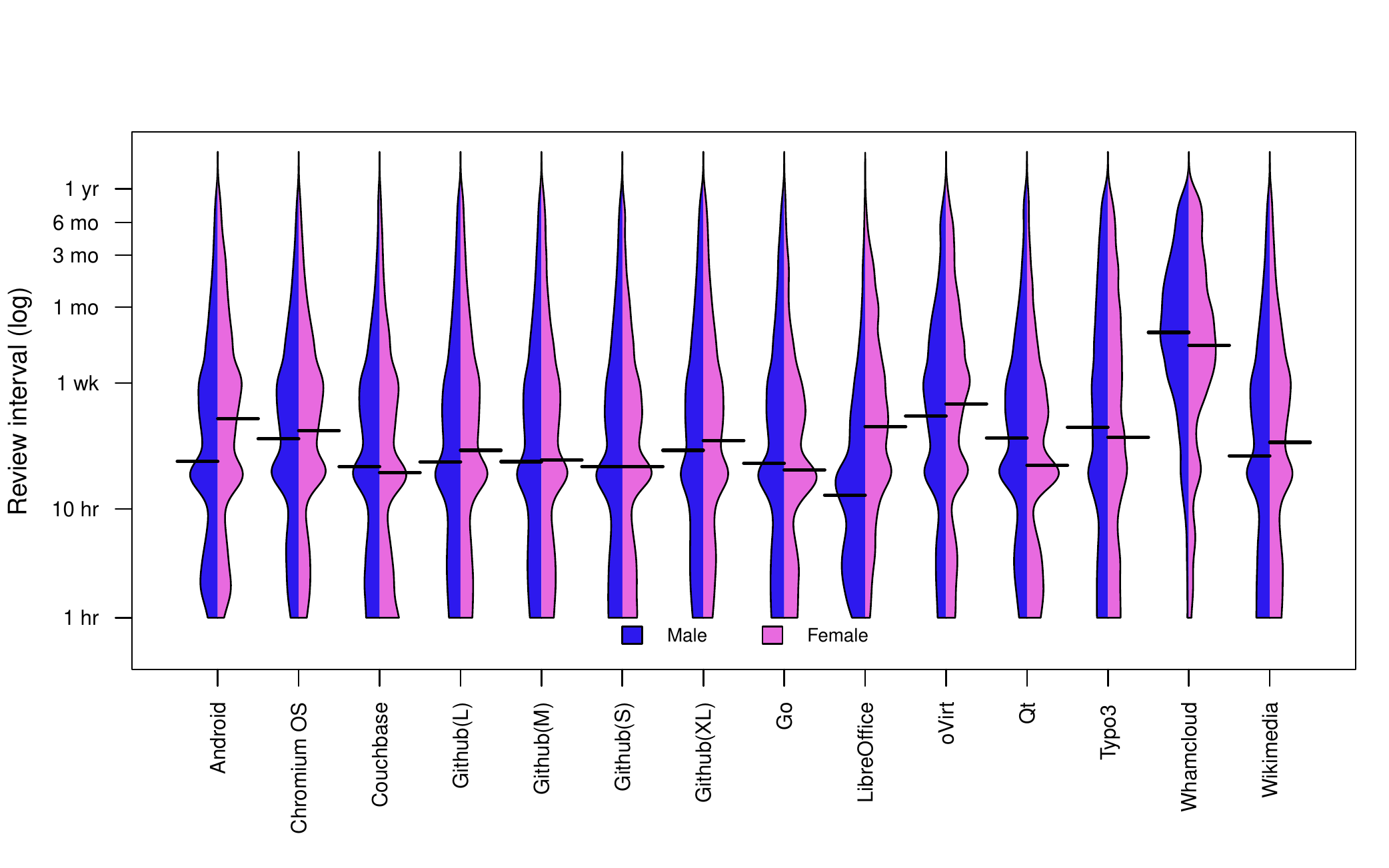}
	\caption{Code review intervals: Men vs. Women. The middle line in a bean plot represents the median review interval for the associated group.}
	\label{fig:thumbnail-ss}	
	\vspace{-12pt}
\end{figure}

 \begin{boxedtext}
\revised{\textbf{RQ2:} \emph{
 Do the genders of the contributors influence code review intervals for their code changes?}}
\end{boxedtext}
  Figure \ref{fig:thumbnail-ss} shows the distribution for review interval hours for men and women using bean plots. We log-transformed review interval hours to account for skewed distributions. 
 Table \ref{table:RQ2} shows correlations of gender and  GNPs with code review intervals based on our linear regression models. 
 Since the dependent variable for these models is log-transformed, exponentiating the coefficients yields the multiplication factor for that variable. For example, in the Android project the coefficient for the \code{Gender} = 0.112. Therefore, when \code{Gender} changes from 0 to 1 (i.e., man to woman), code review intervals increase $e^{0.112} =1.12$ times. In Table~\ref{table:RQ2}, a Delay $>1$ indicates higher code review intervals with the presence of a factor and vice versa.
We also expressed the fit of the model for each project with Adjusted $R^2$. 
 Similar to H1,  we trained a second regression model for each project, where \code{Gender} and \code{isGenderNeutral} is replaced with \code{Gender} * \code{isGenderNeutral}. As men with GIPs are the reference group in these models, the values reported in columns 5-7 of Table~\ref{table:RQ2} indicate a delay multiplier with respect to men with GIPs.  For example, for the Chromium OS project, the delay multiplier for women with  GIPs and women with GNPs are 1.07 and 0.77 respectively, with respect to men with GIPs. 
 
\begin{table*}
	\caption{ Do gender/gender-neutral profiles of the contributors influence code review intervals for their code changes?}
	\centering \label{table:RQ2}
	\resizebox{\linewidth}{!}{%
\begin{tabular}{| l | l | l | l || l | l |l |} 
\hhline{~~-----}
\multicolumn{2}{c}{}  & \multicolumn{5}{|c|}{\textbf{Values indicate \{delay multiplier\}$^{p-value}$ }}\\ \hline
\textbf{Project} & \textbf{Adj. } & \multicolumn{1}{c|}{\textbf{Gender}} & \multicolumn{1}{c||}{\textbf{Neutral}}  & \multicolumn{1}{c|}{\textbf{Women}} & \multicolumn{1}{c|}{\textbf{Men}} & \multicolumn{1}{c|}{\textbf{Women}} \\ 
 &  $R^2$ &  & \textbf{profile}   &   \textbf{with GIP}   &  \textbf{ with GNP}   & \textbf{with GNP}   
     \\ [0.5ex] \hline
  
Android & 0.433 & \biasToMen{1.12}$^{***}$ & {1.04}$^{***}$ & 1.11$^{***}$ & 1.04$^{***}$ & 1.01  \\ \hline 
Chromium OS & 0.235 & {1} & {1.07}$^{***}$ & 1.24$^{***}$ & 1.1$^{***}$ & \biasToGendered{0.77}$^{***}$  \\ \hline 
Couchbase & 0.312 & \biasToWomen{0.85}$^{***}$ & {1.16}$^{***}$ & 2.02$^{***}$ & 1.17$^{***}$ & \biasToGendered{0.39}$^{***}$  \\ \hline 
Go & 0.287 & \biasToWomen{0.73}$^{***}$ & {1.13}$^{***}$ & 0.72$^{***}$ & 1.13$^{***}$ & 1.02  \\ \hline 
 LibreOffice & 0.186 & \biasToMen{2}  $^{***}$ &  $-$ &  $-$ &  $-$  & $-$   \\ \hline 
oVirt & 0.398 & {1.01} & 0.96$^{**}$ & 0.78$^{***}$ & 0.93$^{***}$ & 1.45$^{***}$  \\ \hline 
 Qt & 0.107 & \biasToWomen{0.54}  $^{***}$ &  $-$ &  $-$ &  $-$  & $-$   \\ \hline 
Typo3 & 0.183 & \biasToMen{1.13}$^{*}$ & 0.76$^{***}$ & 1.05 & 0.75$^{***}$ & 1.84$^{***}$  \\ \hline 
 Whamcloud & 0.268 & \biasToWomen{0.77}  $^{***}$ &  $-$ &  $-$ &  $-$  & $-$   \\ \hline 
 Wikimedia & 0.265 & {1.01}   &  $-$ &  $-$ &  $-$  & $-$   \\ \hline 
Github (S) & 0.091 & \biasToWomen{0.8}$^{***}$ & {1.47}$^{***}$ & 0.72$^{***}$ & 1.45$^{***}$ & 1.23$^{*}$  \\ \hline 
Github (M) & 0.092 & \biasToWomen{0.89}$^{***}$ & {1.03}$^{**}$ & 0.93$^{**}$ & 1.04$^{***}$ & \biasToGendered{0.91}$^{**}$  \\ \hline 
Github (L) & 0.099 & \biasToWomen{0.85}$^{***}$ & {1.15}$^{***}$ & 0.74$^{***}$ & 1.12$^{***}$ & 1.39$^{***}$  \\ \hline 
Github (XL) & 0.109 & \biasToWomen{0.97}$^{***}$ & {1.15}$^{***}$ & 0.89$^{***}$ & 1.14$^{***}$ & 1.18$^{***}$  \\ \hline

 \multicolumn{7}{p{11cm}}{Cells in \bluebg{blue} backgrounds represent men with significantly lower review intervals and cells in \pinkbg{pink} backgrounds represent the same for women. For the $p$ value columns, *** , **, and *  represent values $<$ 0.001, $<$ $<$ 0.01, and $<$ 0.5 respectively. No values among $p$ columns indicate statistically insignificance. For the gender neutral hypotheses test (i.e., H2.2), \graybg{gray} background suggests  significant  lower review intervals for women with gender neutral profiles.  }
\end{tabular}}

	\vspace{-8pt}
\end{table*}

\subsubsection{Results of H2.1}
\noindent \emph{Gerrit dataset:}
Seven out of the ten projects show significant differences in CR intervals based on Gender and thereby supporting $H2.1_a$. For the remaining three projects (i.e., Chromium OS, oVirt, and Wikimedia) the null hypotheses ($H2.1_0$) cannot be rejected.
CRs submitted by women encounter significantly higher delays than men among three projects (Android, LibreOffice, and Typo3). On the other hand, the opposite is true for  Couchbase, Go, Qt, and Whamcloud.

\noindent \emph{GitHub dataset:} Our results for all four GitHub datasets support  $H2.1_a$, with women having significantly lower CR intervals than men in all cases.

\subsubsection{Results of H2.2}
\noindent \emph{Gerrit dataset:}
Our results also suggest significant differences between developers with and without GNPs. Four of the projects indicate additional delays for authors with GNPs than authors with GIPs. On the other hand, two projects (i.e., oVirt and Typo3) indicate shorter CR intervals for GNP groups. 
Delay values under \code{Woman with GIP} indicate that women from this group have significantly longer CR intervals than men with GIPs in three projects (i.e., Android, Chromium OS, and Couchbase), while Go and Ovirt show the opposite association. 

All six projects indicate significant differences in CR intervals between men with GIPs and men with GNPs (i.e., supporting $H2.2_a$). For men, GNP is associated with significantly longer delays in four projects and significantly the opposite in the remaining two. Interestingly, although GNP is associated with shorter review intervals for men from oVirt and Typo3, this factor shows the opposite association for women from the same projects.
Comparisons between women with GIPs and women with GNPs suggest significantly lower CR intervals for the GNP group in two projects (i.e., Chromium OS, and Couchbase) and the opposite in two (i.e., Typo3 and oVirt).  Therefore, $H2.2_a$ is supported for women from four projects.
In summary, all six projects support $H2.2_a$. 
with either men or women showing significant differences in CR intervals based on GIP selection.

\noindent \emph{GitHub dataset:}
 We also noticed significant associations between GNP and CR intervals among all four GitHub groups.
 Participants with GNPs were more likely to encounter longer CR intervals than participants with GIPs.  
 Similarly, women with GNPs had significantly longer CR intervals than men with GIPs among three of the four GitHub groups.  The remaining GitHub group (i.e., M)  shows significantly shorter CR intervals for such women.
 
 Men with GNPs had significantly longer CR intervals than men with GIPs in all four groups (i.e., support for $H2.2_a$).  
  Comparisons between women with GIPs and GNPs  suggest significantly longer CR intervals  for the GNP group in all cases except M. In aggregate, 
$H2.2_a$ is supported by all four GitHub groups.

 \begin{boxedtext}
 \emph{
 Our results support $H2.1_a$ for seven out of the 10 Gerrit projects. However, we fail to reject the null hypothesis $H2.1_0$ for three such projects (i.e.,
Typo3, oVirt, and Wikimedia). We found support for $H2.2_a$ in all six Gerrit-based projects allowing GIPs. We also found support for both $H2.1_a$ and
$H2.2_a$ across all four GitHub project groups. GNPs are associated with
significantly longer code review intervals in eight out of the 10 cases (i.e.,
6 Gerrit-based, and 4 GitHub groups). However, women with GNPs had significantly shorter CR intervals in three of the cases
– Chromium OS, Couchbase, and GitHub (M).
}
\end{boxedtext}

\subsection{Gender vs. code review participation (RQ3)}

\begin{boxedtext}
\revised{\textbf{RQ3:} \emph{
 Do the genders of the contributors influence their participation in code reviews?}}
\end{boxedtext}

\begin{figure}
	\centering  \includegraphics[width=\linewidth]{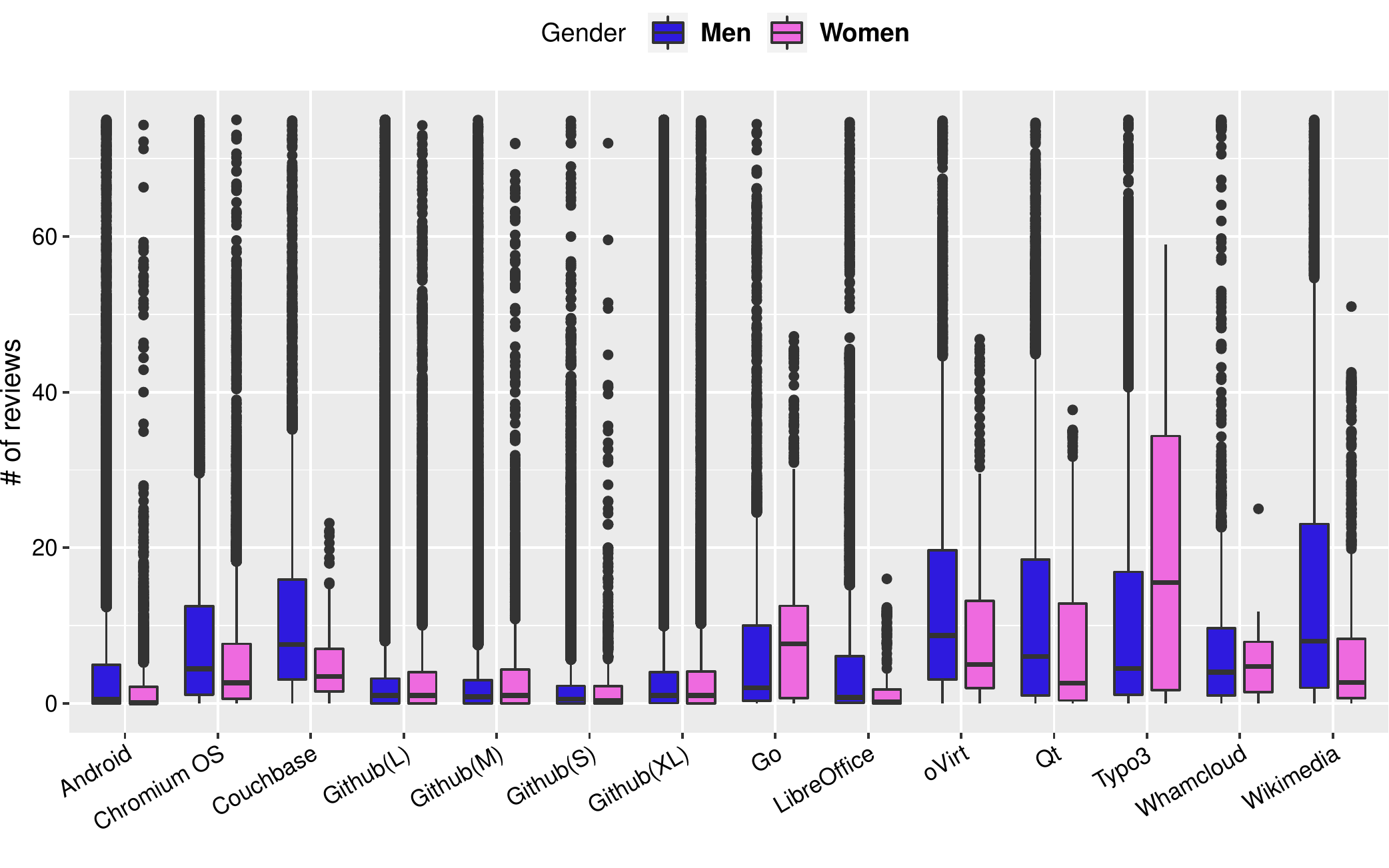}
	\caption{Distribution of review counts per month for men and women}
	\label{fig:thumbnail-ss-2}	
	\vspace{-12pt}
\end{figure}
\begin{table*}
	\caption{Do gender/gender-neutral profiles of the contributors influence their participation in code reviews?}
	\centering \label{table:RQ3}
	\resizebox{\linewidth}{!}{%
\begin{tabular}{| l | l | l | l || l | l |l |} 
\hhline{~~-----}
\multicolumn{2}{c}{}  &\multicolumn{5}{|c|}{\textbf{Values indicate \{regression coefficient\}$^{p-value}$ }}\\ \hline
\textbf{Project} & \textbf{Adj. } & \multicolumn{1}{c|}{\textbf{Gender}} & \multicolumn{1}{c||}{\textbf{Neutral}}  & \multicolumn{1}{c|}{\textbf{Women}} & \multicolumn{1}{c|}{\textbf{Men}} & \multicolumn{1}{c|}{\textbf{Women}} \\
 &  $R^2$ &  & \textbf{profile }   &   \textbf{with GIP}   &  \textbf{ with GNP}   & \textbf{with GNP}   
     \\ [0.5ex] 
 \hline
Android & 0.244 & \biasToMen{-1.6}$^{***}$ & 0.58$^{***}$ & -1.64$^{***}$ & 0.57$^{***}$ & 0.07  \\ \hline 
Chromium OS & 0.307 & \biasToMen{-2.14}$^{***}$ & -0.05 & -3.43$^{***}$ & -0.23 & \biasToGendered{1.57}$^{***}$  \\ \hline 
Couchbase & 0.386 & \biasToMen{-3.17}$^{***}$ & {1.81}$^{***}$ & 5.14$^{***}$ & 2.23$^{***}$ & -9.49$^{***}$  \\ \hline 
Go & 0.419 & \biasToMen{-1.63}$^{*}$ & -1.06$^{**}$ & -5.27$^{**}$ & -1.26$^{**}$ & \biasToGendered{4.35}$^{*}$  \\ \hline 
 LibreOffice & 0.343 & -0.85   &  $-$ &  $-$ &  $-$  & $-$   \\ \hline 
oVirt & 0.429 & \biasToMen{-3.5}$^{***}$ & -0.8$^{*}$ & -14.05$^{***}$ & -2.25$^{***}$ & \biasToGendered{13.7}$^{***}$  \\ \hline 
 Qt & 0.574 & \biasToMen{-1.77}  $^{*}$ &  $-$ &  $-$ &  $-$  & $-$   \\ \hline 
Typo3 & 0.273 & \biasToWomen{4.57}$^{***}$ & 0.6 & 6.2$^{***}$ & 0.86 & -5.69$^{*}$  \\ \hline 
 Whamcloud & 0.224 & \biasToMen{-2.15}  $^{*}$ &  $-$ &  $-$ &  $-$  & $-$   \\ \hline 
 Wikimedia & 0.504 & \biasToMen{-4.23}  $^{***}$ &  $-$ &  $-$ &  $-$  & $-$   \\ \hline 
Github (S) & 0.078 & \biasToMen{-0.63}$^{*}$ & 0.62$^{***}$ & -0.73 & 0.61$^{***}$ & 0.2  \\ \hline 
Github (M) & 0.094 & -0.05 & -0.42$^{**}$ & 0.86$^{*}$ & -0.25 & -1.92$^{***}$  \\ \hline 
Github (L) & 0.171 & \biasToMen{-0.7}$^{***}$ & -0.45$^{***}$ & -1.22$^{***}$ & -0.54$^{***}$ & \biasToGendered{1.14}$^{***}$  \\ \hline 
Github (XL) & 0.138 & \biasToMen{-1.25}$^{***}$ & -0.84$^{***}$ & -1.61$^{***}$ & -0.89$^{***}$ & \biasToGendered{0.68}$^{*}$  \\ \hline 

   \multicolumn{7}{p{12cm}}{Cells in \bluebg{blue} backgrounds represent men with higher code review participation and cells in \pinkbg{pink} backgrounds represent the same for women. For the $p|$ value columns, *** , **, and *  represent values $<$ 0.001, $<$ $<$ 0.01, and $<$ 0.5 respectively. No values among $p$ columns indicate statistically insignificance. For the gender neutral hypotheses test (i.e., H3.2), \graybg{gray} background suggests  a significant higher number of code review participation from women with neutral profiles.  }
\end{tabular}
}
	\vspace{-8pt}
\end{table*}

Figure \ref{fig:thumbnail-ss-2} shows the distribution of CR participation for men and women. The long-tailed distributions indicate that some of the individuals take significantly higher review loads than usual during some monthly intervals.
Figure~\ref{fig:thumbnail-ss-2} also indicates that median review participation from women is lower than men in 12 out of the fourteen cases, where  Go and Typo3 are exceptions. 

 Table \ref{table:RQ2} shows correlations of gender and  GNPs with CR participation based on our linear regression models. Column Adj. $R_2$ reports the goodness of fit for the trained models.
For these models, linear regression coefficients represent how CR participation per month changes if a factor changes from 0 to 1.  A positive coefficient ($Z$) value indicates a positive correlation with a factor and vice versa. For example, in the Android project, the coefficient for the \code{Gender} is -1.60. Therefore, if we change \code{Gender} from 0 to 1 (i.e., man to woman) while keeping all other factors the same, monthly CR participation reduces by 1.60.
Similar to H1 and H2,  we trained a second regression model for each project by replacing \code{Gender} and \code{isGenderNeutral}  with the interaction term, \code{Gender} * \code{isGenderNeutral}.
 
\subsubsection{Results of H3.1}
\noindent \emph{Gerrit dataset:} Nine out of the ten projects indicate significant gender biases in CR participation and therefore support $H3.1_a$.  LibreOffice is the only exception, where the null hypotheses ($H3.1_0$ cannot be rejected.  Eight out of those nine projects indicate women review significantly lower numbers of CRs than men from the same project.   Women had significantly higher CR participation only in Typo3.

\noindent \emph{GitHub dataset:} We observe that men review in significantly higher numbers of PRs than women in three out of the four groups.  Only the M group is an exception, where the difference is not statistically significant. Therefore, $H3.1_a$ is supported for three groups and the null hypothesis ($H3.1_0$) cannot be rejected for the remaining one.

\subsubsection{Results of H3.2}

\noindent \emph{Gerrit dataset:} Our results also indicate a significant association between CR participation and GNPs in four out of the six custom avatar-supporting projects.  While GNP has a positive association with CR participation in Android and Couchbase, the opposite association is seen in  Go and oVirt. 
Four out of the six projects also suggest that women with GIPs have significantly lower CR participation than men with GIPs belonging to the same project. The remaining two projects (i.e., Couchbase, and Typo3) indicate the opposite, where women with GIPs had significantly higher CR participation. 

Four out of the six projects, excluding Chromium OS and Typo3, indicate significant differences in CR participation between men with GIPs and men with GNPs (i.e., supporting $H3.2_a$). While GNP is associated with lower CR participation for men in  Go and oVirt,  opposite associations are found in Android and Couchbase. 
Comparisons between women with GIPs and women with GNPs indicate significant differences in CR participation (i.e., support for $H3.2_a$) among five projects, where Android is the only exception. Women with GNPs have significantly higher CR participation than women with GIPs in three projects (i.e., Chromium OS, Go, and oVirt), and the remaining two projects ( i.e., Couchbase and Typo3) indicate the opposite. In summary, all six projects support $H3.2_a$ with either men or women showing significant differences in CR participation based on GIP selection.

\noindent \emph{GitHub dataset:} 

Three GitHub groups except the S group indicate that participants with GNPs were more likely to participate in a significantly lower number of code reviews than participants with GIPs.   
While women with GIPs had significantly lower CR participation than men with GIPs in two groups (i.e., L and XL, the opposite trend is seen in the S group. 

Men with GIPs participate in significantly higher numbers of CRs than men with GNPs among projects belonging to L and XL groups, while projects belonging to S show an opposite trend. In contrast, women with GIPs had significantly lower CR participation than women with GNPs in L and XL groups, while the M group showed an opposite association. In summary, all four GitHub groups show significant differences in CR participation between men/women with and without gender-identifiable profiles and therefore support $H3.2_a$. 

 \begin{boxedtext}
 \emph{Our analyses found support for $H3.1_a$ in nine out of the ten Gerrit-based projects with Libreoffice being an exception.  Three GitHub groups excluding the M group also support $H3.1_a$.  All six GIP-supporting Gerrit projects as well as all four GitHub groups also support $H3.2_a$. }
\end{boxedtext}

\subsection{Results summary}

\begin{table*}
	\caption{Summary of the gender biases among 10 Gerrit datasets and 4 Github datasets. The numbers under a gender indicates the number of groups had biases favoring that gender.}
	\centering \label{table:gender-summary}
	\resizebox{\linewidth}{!}{%

\begin{tabular}{| l | r | r | r | r | r |r |l|} 
\hline
\multirow{2}{*}{\textbf{Hypothesis}}  & \multicolumn{3}{c|}{ \textbf{Gerrit favoring }}   & \multicolumn{3}{c|}{ \textbf{Github favoring }} \\ \hhline{~------}
&  \# biases &\biasToMen{ Men}  & \biasToWomen{Women}  & \#total & \biasToMen{ Men}  & \biasToWomen{Women} \\\hline

\textbf{Code acceptance: \boldmath{$H1.1_a$}}   & 9  & 7 & 2 & 4 & 0 & 4 \\ \hline
\textbf{Review interval: \boldmath{$H2.1_a$}} & 7 & 3 & 4 & 4 & 0& 4 \\ \hline 
\textbf{Review participation: \boldmath{$H3.1_a$}} & 9 & 8 &1  & 3 & 3 & 0\\ \hline
\end{tabular}
}
	\vspace{-8pt}
\end{table*}

\begin{table*}
	\caption{Summary of the biases due to gender-neutral profiles among 6 Gerrit datasets and 4 Github datasets. The numbers under a profile type indicates the number of cases with biases favoring that group.}
	\centering \label{table:neutral-summary}
	\resizebox{\linewidth}{!}{%

\begin{tabular}{| l | r | r | r | r | r |r |l|} 
\hline
\multirow{2}{*}{\textbf{Hypothesis}}  & \multicolumn{3}{c|}{ \textbf{Gerrit favoring }}   & \multicolumn{3}{c|}{ \textbf{Github favoring }} \\ \hhline{~------}
&  \# biases & { GIP}  & \biasToGendered{GNP}  & \#total & { GIP}  & \biasToGendered{GNP} \\\hline

\textbf{Code acceptance: \boldmath{$H1.2_a$}}   & 4  & 3 & 1 & 4 & 2 & 2 \\ \hline
\textbf{Review interval: \boldmath{$H2.2_a$}} & 6 & 4 & 2  & 4 & 4&  0 \\ \hline 
\textbf{Review participation: \boldmath{$H3.2_a$}} & 4 & 2 & 2  & 4 & 3 & 1\\ \hline
\end{tabular}
}
	\vspace{-8pt}
\end{table*}

\revised{Table \ref{table:gender-summary} provides an overview of the gender biases identified among our datasets. According to these results, gender biases are common during CRs/PRs. Review interval (i.e., H2) is the metric, which shows the lowest number of gender biased cases. Even so, 11 out of the 14 cases show biases in terms of review intervals. However, the directions of biases vary based on both the metric and project. In general, among the the Gerrit projects men have favorable measures among majority of the cases. Interestingly, women on Github have favorable measures in terms of both code acceptance and review interval. However, we find an opposite picture for Github women in terms of code review participation.}

\revised{Table \ref{table:neutral-summary} summarizes the biases due to gender neutrality of  profiles. These types of biases are also  present among most of the cases.  Although, persons with GIPs are more likely to have favorable measures, GNP favoring projects are also common.}

\section{Discussion}
\label{sec:discussion}

\textbf{Comparison with Terrell \textit{et} al.  \cite{terrell_1}: }

We investigated two key results from Terrell \textit{et} al.'s study regarding patch acceptance and appearance as women.
First, supporting their findings, we found significantly higher odds of PR acceptance for women on GitHub across all four groups. However, the same finding applies to only two out of the ten Gerrit-based projects, while seven projects suggest the contrary. 

Second, we found three Gerrit-based projects (i.e., Chromium OS, Couchbase, and Go), where women with GNPs have significantly higher odds of getting code accepted than women with GIPs. This finding is similar to Terrell \textit{et} al.'s. However, contrasting theirs, our findings suggest that GNPs significantly lower the odds of acceptance for women's code on GitHub as well as two Gerrit-based projects (i.e., Android, and oVirt). However, we would like to restate that these differences may be due to a different heuristic adopted by this study than Terrell et al.'s to identify GNPs.

\vspace{2pt} \noindent
\textbf{Comparison with Bosu and Sultana~\cite{bosu-esem-2019}: } 
Bosu \etal~\cite{bosu-esem-2019} and our study have nine common projects and our study partially replicates their study, some of our results differ from their findings. 
First, they found less than 10\% women across all the projects. However, we noticed four projects, including three common to both studies, exceeding that threshold.

Their results found significantly lower acceptance rates for women in Android, Chromium OS, and LibreOffice and the opposite in oVirt, Qt, and Typo3. 
While we obtained similar results for Chromium OS, and LibreOffice, where women had lower odds of acceptance, we notice the opposite in Android. 
Similarly, although our results for Qt are similar to Bosu and Sultana~\cite{bosu-esem-2019}, the opposite is seen for oVirt. Our results of Typo3 are inconclusive. 

For our second hypothesis ($H2.1$), Bosu and Sultana~\cite{bosu-esem-2019} found biases favoring men in Android and Couchbase whereas women are favored in Qt and Typo3. On the contrary, we found two projects (i.e., Couchbase and Typo3) flipping with women being favored in Android and Typo3 but men being favored in Couchbase and Qt. 
However, these differences may be due to two study factors. First, we have added four additional years of data. Second, unlike Bosu and Sultana, we have used regression models to account for various confounding variables. 
 Regardless, we comply with Bosu and Sultana's hypotheses~\cite{bosu-esem-2019} that gender biases are not universal and do not work in the same direction across all projects.  The prevalence of gender bias may depend on various contextual factors, such as the culture of a community, sponsor, governance, and project size.

\vspace{2pt} \noindent
\textbf{Women are more likely to use gender neutral profiles: } Table \ref{table:gender_demographics} shows that  majority of women (52\%) prefer GNPs. 
On Gerrit-based projects more than three-quarters of women use GNPs. While the ratio of women using GNPs is lower on GitHub than women on Gerrit, almost half of the women on GitHub use GNPs. In comparison, almost two third men use GIPs on GitHub. 
Prior studies have found women using GNPs to avoid unwanted attention~\cite{frluckaj2022gender,menking2019people} as well as avoid biases~\cite{terrell_1}. 
While using GNP is a personal choice for a person, disproportional higher ratios of  GNPs among women hinder promoting diversity and inclusion initiatives, as prior studies have found a lack of mentors and women role models as a barrier to encouraging more women joining computing~\cite{bosu-esem-2019,wang2020reasons,lockwood2006someone}.

\vspace{2pt} \noindent
\textbf{Do gender-neutral profiles help avoid biases?}
Terrell \textit{et} al. hypothesized that GNPs may help women avoid unwanted attention as well as biases on GitHub. While we found evidence supporting this hypothesis in three Gerrit-based projects (i.e., Chromium OS, Couchbase, and Go), we also found contradictory evidence in two projects and among all four GitHub groups. 
Women with GNPs had between 1-5\% lower PR acceptance rates than women for GIPs from the same GitHub group. 
On the contrary, men with GNPs had significantly higher odds of acceptance than men with GIPs in two GitHub groups (i.e., M, and L). Therefore, GNP may not be a consequential factor, especially on GitHub. 
However, GNPs may be helpful for women in some projects. For example, women with GNPs from the Go project had  +16.1\% higher acceptance rates than women with GIPs from the same project. On the contrary, the difference in acceptance rate between Go project's men with and without GNPs is -5.55\%.  Therefore, while GNPs are positively associated with women's acceptance of Go, the opposite association is found for men. 
Regardless,  GNPs do not solve underlying biases and help promote diversity and inclusion. Therefore, if a project manager finds similar differences in another project, they should investigate possible causes and their remedies.

\vspace{2pt} \noindent
\textbf{Gender bias exists in different directions:} From the results of three research questions, we see that gender bias exists in FOSS communities but the direction of bias varies. In terms of code acceptance, we found seven Gerrit projects significantly favoring men but Android, Qt, and four GitHub groups favored women. In terms of CR intervals, men are favored among three Gerrit projects, and women are favored in four Gerrit projects as well as the four GitHub groups.  Finally, in terms of CR participation men had even large advantages with eight Gerrit projects as well as three GitHub groups in their favor. 
Therefore, gender biases may manifest not only in different directions in another project but also differently across other CR aspects in the same project. \textit{One size fits all} approach may not be the best strategy to mitigate gender biases during CRs.

\vspace{2pt} \noindent
\textbf{Women participate less in code review:} 
Women encounter the most biases during CR participation with eight Gerrit projects and three GitHub groups being against their favor. 
While a person may self-assign to participate in a CR, most of the CR assignments are based on invitations. Therefore, women are less likely to be invited to participate than men. Since lower CR participation from women is a trend common across almost all the projects, this observation may be explained by affinity bias --the tendency of people to collaborate with others who share similar interests, experiences, and backgrounds. Moreover, a recent study also suggests affinity biases during pull request acceptance on GitHub~\cite{nadri2021relationship}. 

We also noticed another interesting trend among several projects. Although men with GNPs from  Chromium OS, Go, oVirt, GitHub (L), and GitHub (XL)   are less likely to participate in CRs than men with GIPs, we see the opposite trend for women with GNPs having higher participation.  This observation further supports our hypotheses regarding affinity biases. 
Since CRs have other benefits such as knowledge sharing and building relationships~\cite{bosu2016process}, lower CR participation puts women in disadvantageous positions to men counterparts. Therefore, to promote diversity and inclusion, FOSS projects should take initiatives to ensure equitable participation of women in the CR process. 

\begin{figure}
	\includegraphics[width=\linewidth]{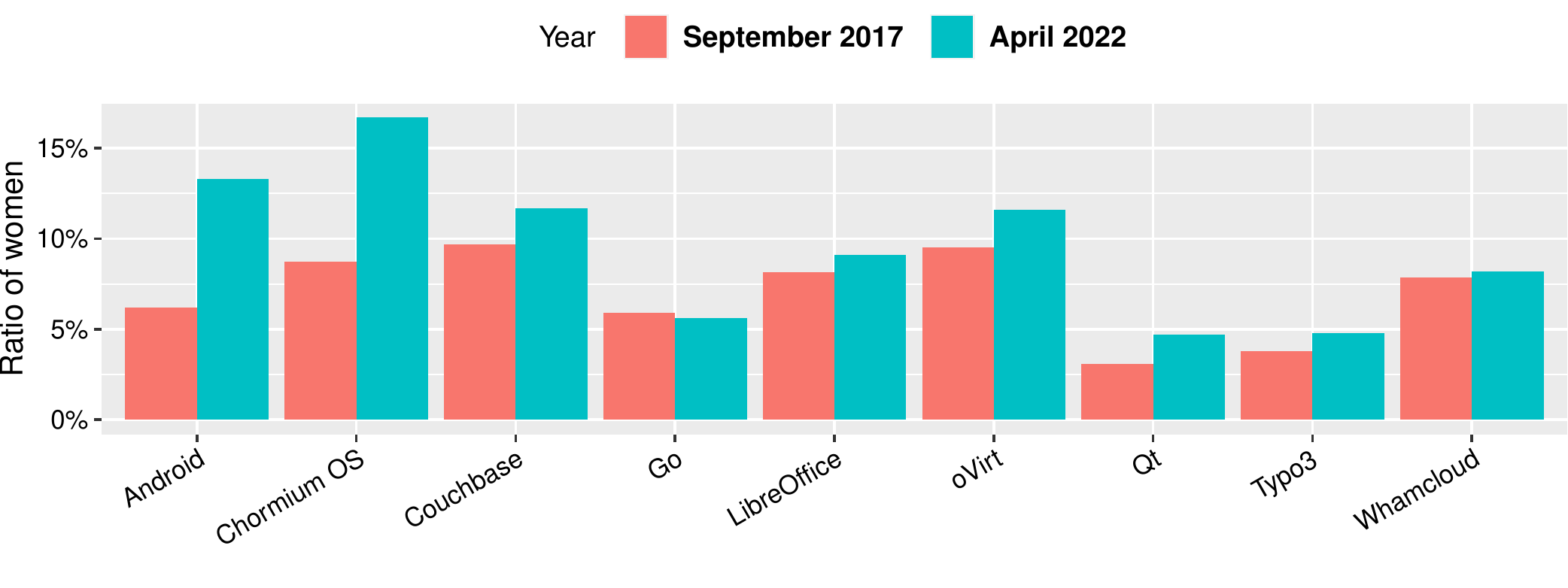}
	\caption{Ratio of women in 2022 with comparison to  2017}
	\label{fig:women-ratio}	
	\vspace{-12pt}
\end{figure}

\vspace{2pt} \noindent
\textbf{Ratio of women increasing among FOSS projects:} Since nine out of the ten projects were also included in Bosu and Sultana's study~\cite{bosu-esem-2019}, we compare ratios of women reported for each of these projects in their study against those found in this study (Figure~\ref{fig:women-ratio}). Eight out of the nine projects indicate higher ratios of women in April 2022 than in September 2017, where Go is the only exception. Android and Chromium OS indicate the highest increment.

\vspace{2pt} \noindent
\textbf{Recommendations for project managers:} \revised{Project managers can play an important role in fostering diversity and inclusion. Many organizations use platforms and dashboards to improve software development practices including code reviews~\cite{hasan2021using,bird2015lessons,augustine2017deploying,guerrouj2016software}. These platforms can provide them with insights into code review activities and uncover potential biases. For example, if a woman is among the top contributors based on code commits but ranks way lower in terms of code review participation,  or if a woman is considered an expert for a project area but is not added as a reviewer for relevant changes, those may hint potential biases against them. Project managers should look into such cases and delve further to identify possible causes. If biases are found, a manager should take steps such as educating developers, arranging team-building activities,  or creating policies to mitigate biases.}

\vspace{2pt} \noindent
\textbf{Directions for researchers:} Diversity and impact of gender bias in FOSS projects are well researched. Researchers also suggested numerous strategies to mitigate such biases. Though gender bias exists among many projects, the direction and amplitude of bias vary based on project size, community, and culture. Similar strategies may not work across all projects, characteristics of biases and their underlying causes differ. While prior research has proposed various bias mitigation strategies (detailed in Section~\ref{subsec:bias-strategy}), how various contextual factors influence each strategy's relevance as well as effectiveness, lacks empirical validation. 

Women are subject to not only biased feedback during code reviews~\cite{paul_sentiment, nasif_icse}, but also less likely to have equitable participation in the CR process.
Results of our study suggest CR participation as one of the most gender inequitable aspects among most of the projects. Therefore, future work should focus on the identification of barriers to promote equitable participation of women in the CR process, and strategies to mitigate those barriers.

\section{Related works}
\label{sec:related-works}

This section briefly presents existing research on gender bias in open-source communities.

\subsection{Characteristics of gender bias in FOSS}

Several prior studies have found the existence of gender bias in FOSS projects~\cite{terrell_1,bosu-esem-2019}. Bosu and Sultana found that women consist only 6.70\% of the non-casual developers in 10 popular Gerrit projects. While Vascilescu et al. found 9\% women among approximately 873k GitHub contributors~\cite{vasilescu2015gender}, in a later study Terrell et al. found only 5.2\% among 4 million GitHub users~\cite{terrell_1}. These studies suggest gender unbalance among FOSS communities. Wajcman shows a connection between masculinity and technology and suggests that this connection results from the historical and cultural construction of gender and produces gender bias in tech~\cite{wajcman2007women}. Biases preferring one gender over another can be explicit or implicit. Prior research shows that implicit gender biases are dominant in professional software developers and impact hiring decisions and contribution evaluation. Bosu and Sultana~\cite{bosu-esem-2019} found explicit biases against women in the form of lower code acceptance rate, longer first feedback interval, longer code review interval, and higher scrutiny during code review. They also suggest the existence of implicit biases regarding participation in code review.

Women also encounter bias during communicating with other contributors. Nafus~\cite{nafus_2012} found that men monopolize authorship of code in the FOSS community and withdraw necessary social ties that are necessary to build a gender-inclusive environment. Canodo \textit{et} al.~\cite{breaking_one_barrier} found that lack of gender-equal communication impedes the participation of women. Women participants from the study of Lee and Carver~\cite{lee_carver} reported negative interactions from males towards females. Also, women shared that men are aggressive and hold stereotyping beliefs like women are less competent, devalue women's input, and try to humiliate women during interactions. Moreover, while investigating how FOSS contributors interact with opposite genders they found both positive and negative responses. Negative responses express ideas related to stereotypes about women, like women getting easily offended or creating drama. One respondent shared that \textit{men are usually more used to being rejected, so there is less potential drama to be afraid of} about interacting with women contributors. On the contrary, female contributors shared that they were required to prove themselves repeatedly or were made to feel less comfortable around men.  As contemporary software development organizations are dominated by males, occurrences of misogynistic and sexist remarks are abundant in many communities~\cite{squire2015floss}. 

Along with community and organizations software products or tools can be biased to one group of people neglecting others \cite{gendermag_1}. Gender inclusiveness issues have also been found in software tools. Researchers have developed a tool GenderMag~\cite{burnett2016gendermag} to identify gender bias in software tools. Mendez \textit{et} al.~\cite{anita_1} found gender issues in from 7\%-71\% of the use cases for tools and infrastructure used in OSS projects and showed that because of biased infrastructure women newcomers are more disadvantaged than men. It is proven that tools are a contributing factor in gender disparity in the open-source community \cite{anita_1, anita_2} so it is high time we should focus on building gender bias-free tools for software development.

\subsection{Consequences of gender bias}
Vandana \textit{et.} al~\cite{vandana_4} investigated the situation of women working in OSS communities and shared that due to the very low representation of women, they feel isolated and burdened to represent all women who are being excluded. Moreover, women often feel their contributions are not valued enough because of their gender and they have to work harder to get recognition. Surprisingly, one participant in her study shares that women are expected to write better code than men because they have to be outstanding to reach a higher level of projects whereas men with average skillsets can make it to those positions. Nasif and his colleagues~\cite{nasif_icse} conducted a study to investigate the effect of Prove-it-again where a member of a non-stereotypical group has to meet stricter standards to prove his/her competence. They did not find that women's pull requests take longer time, in general, to get accepted but women concentrate their contribution on fewer projects and fewer organizations than men.

According to Edna \textit{et} al.~\cite{breaking_one_barrier}, lack of gender-equal communication hinders the participation of women in OSS communities. In this case, some men do not interact or talk with their female colleagues equally. To express opinions, women have to interrupt others, even if they are project managers. Paul \textit{et} al.~\cite{paul_sentiment} researched sentiment during code reviews for 6 popular OSS projects( Android, Chromium OS, Couchbase, OmapZoom, Ovirt, Qt) and found that in three of these projects, women receive negative reviews more often. Moreover, men withhold positive encouragement while reviewing women's codes. Women tend to express their sentiments less than men and also are more neutral to men counterparts than other women contributors. Additionally, the interaction between men and women was found to be flirtatious whereas women-to-women communication was found supportive\cite{thelwall2010data} in MySpace comments. Nasif and his colleagues~\cite{nasif_icse} also investigated if women have a narrow band of socially acceptable behavior while contributing to OSS projects. They shared that women avoid using profane words more than men, they are more neutral while showing sentiments or stereotypical masculine or feminine traits.

\subsection{Strategies to mitigate bias}
\label{subsec:bias-strategy}
Researchers also mentioned multiple initiatives e.g. promoting women to leadership roles, adopting codes of conduct (CoC) and maintaining transparent organizational culture, promoting gender-inclusive language, de-biasing tools and so on which can be taken to mitigate bias in the OSS platform, and increasing participation of women. Canedo \textit{et} al.~\cite{canedo_2} and Prana \textit{et} al.~\cite{prana} suggest assigning women to senior roles to lead them to empowerment. Leadership increases the visibility of their accomplishments, shows respect, and encourages others to join the team\cite{vandana_2}. Catolino \textit{et} al.~\cite{catalino} shows in their study presence of women in the community reduces community smells that are sub-optimal patterns in the culture of software organization by mediating discussions and increasing communications among members. Avoiding segregation of tasks as masculine and feminine, treating women developers as developers can dismantle stereotypical views and create a gender-inclusive environment \cite{calvo,setting_quota,canedo_2}. Participants of Vandana's study~\cite{vandana_2}  suggested including women in different conferences and activities instead of isolating them. Singh mentions letting women lead where it is appropriate \cite{vandana_2},  

CoC should enlist norms of unacceptable behavior in the community and the consequences of violations of those codes. Singh \textit{et} al.~\cite{vandana_oss} investigated websites for 355 projects in OSS where only 28 of them had Codes of Conduct or similar guidelines. Prior studies show that introducing a code of conduct can reduce tightrope effects and foster a culture of collaboration \cite{nasif_icse,vandana_2,prana,calvo,vandana_oss}. A study conducted by Robson \cite{Robson} shows that it is not enough only to introduce the code of conduct in the community, proper enforcement is also required to make the environment inclusive for all \cite{vandana_2,code_of_conduct_in_OSS}. Moderators should invigilate whether the rules are being followed and punishment must be ensured if necessary \cite{some_of_all}. If organizations that host the open source project remain transparent about their work environment and culture, women can decide beforehand to start a contribution. It can increase the level of affinity between personal and professional values\cite{qiu_stewart}.

Women newcomers in OSS face disproportionately more barriers than others. Among different types of biases, gender bias in tools and infrastructures used in open-source medium disadvantages women \cite{anita_1}. Researchers have developed a tool: GenderMag\footnote{https://gendermag.org/} that can be used to de-bias tools. GenderMag uses gendered personas and specialized Cognitive Walkthrough (CW) to evaluate software and identify if there is any gender-inclusiveness issue for the tool for a wide range of people from designers to software developers \cite{anita_1,anita_2,anita_3}.

Researchers have also mentioned a few humanitarian open-source projects where women might be interested in \cite{hfoss}. Women developers can start contributing to these projects and gain confidence and experience. Such types of projects can be launched in a greater number to attract more women to OSS. In a nutshell, researchers suggested inaugurating projects that are specifically preferred by women.

Qiu \textit{et} al.~\cite{qiu_stewart} investigated the signals which contributors notice before starting to work on a project. Few participants shared that openness of the community or language used by the contributors plays an important role in deciding to contribute to a project. One participant in their study found that one project included ``nice guys" in their documentation. OSS contributors also raised concern about gender-inclusive language sharing that project should avoid words like ``guys" or gendered pronouns, assuming contributors are mostly men or hold one demographic background \cite{canedo_2,qiu_stewart}, Nafus \cite{nafus_2012} also indicated use of inadequate language in OSS. So, it is mandatory to use gender-inclusive language in documentation or during communication with each other.

\section{Validity Threats}
\label{sec:threats}

\textbf{Internal validity:}
We examined our hypotheses for a total of 1010  FOSS projects, where 1000 projects are using pull-request-based code reviews and the remaining 10 are using Gerrit-based code reviews. Project selection is the primary threat to the internal validity of our study. Since we partially replicate the studies of Bosu and Sultana's~\cite{bosu-esem-2019} study, we adopted their project list for the Gerrit dataset. As one of those projects, OmapZoom was inactive, we replaced that with another project Wikimedia that gets along with their project selection criteria. We also updated their dataset with code reviews until 30th April 2022 so that we can get the most recent status of gender bias. 

Furthermore, we are replicating the study of Terrell \textit{et.} al~\cite{terrell_1}. As their dataset is not publicly available, we selected 1000 projects from GHTorrent exported in March 2021. Being an open-source platform GitHub hosts different types of personal, and group projects that may or may not be based on programming languages. Following Kalliamvakou \textit{et} al.'s recommendation,~\cite{kalliamvakou2016depth}, we selected only those projects that use one of the popular programming languages, have at least 20 contributors and have 20 pull requests during the last three months. We followed the criteria to avoid any discontinued or too small projects that may not add any valuable information to our study. Moreover, we selected GitHub projects by stratified sampling so that we can have insights about projects of different sizes and conducted data analysis separately. 
Despite our carefully designed sampling strategy, the characteristics of our sample may differ from the entire GitHub ecosystem. 

GHTorrent 2019 dataset had a data hole for July to December 2019. Though GHTorrent did not confirm if the issue has been fixed  GHTorrent MySQL export from March 2021, we found pull requests created during that period in this dataset. However, we cannot confirm if all the pull requests during that period are included in our analyses. 

\vspace{2pt} \noindent \textbf{External validity:} We can observe from the results that direction and degree of bias differ for different projects. Results might be different if we replicate the study in different settings or different projects. The scenario for communication and collaboration among developers may not be the same for other projects in FOSS communities. A project's culture also depends on various factors such as the nature and size of the projects, the number of contributors, and project governance. While we have used data from a large number of projects in this study, our sample may not adequately represent the entire FOSS spectrum.

\vspace{2pt} \noindent \textbf{Construct validity:} One major construct threat to the validity of our study is the gender resolution of the users. 
First, we did not attempt to identify non-binary genders, since it is almost impossible to identify those developers without their input.
Second, for the persons with GNPs, we assign GenderComputer's resolution. While GenderComputer has been used in several prior SE studies including the two original ones, name-to-gender resolution tools are prone to misclassifications~\cite{santamaria2018comparison}. Therefore, between 5-15\% of the 85,772 persons with GNPs are miscategorized in our dataset. 
Finally, for the users with GIPs, we assigned genders based on their avatars. If we identified a conflict between the gender assigned by GenderComputer and our avatar-based assignment, we manually checked the profile and searched on social media to make a resolution. However, if GenderComputer assigns either `Unisex' or `None' and a user is using an avatar suggestive of a different gender, he/she would be mislabeled.  While we believe such cases would be negligible, such misclassifications are another threat to our gender resolution strategy. 

We used python library \textit{lizard} to calculate source code metrics, which have been starred more than a thousand times on GitHub and have been forked by more than 200 times. While we have not evaluated its accuracy, it may be subject to some errors.

\vspace{2pt} \noindent \textbf{Conclusion validity:} 
We have regression-based models for analysis to account for several confounding variables. We have used well-known and matured libraries such as \code{stats}, \code{rms} to train our models. To estimate effects, we have used standard metrics such as the Odds ratio for logistic regression models and co-efficient ($Z$) for the linear models. Therefore, it is unlikely to arise any threats to validity from the evaluation metrics, library selection, or evaluation of the dependent variables.

\section{Conclusion}
\label{conclusion}
Gender bias in FOSS communities is well-researched. But how this bias varies for different aspects of OSS project development is not studied. To bridge this gap, we studied the existence of bias in both explicit (pull request acceptance, code review interval) and implicit bias ( code review participation). We found that gender bias exists in different directions. The level of bias in the above-mentioned aspects differs based on the size of the projects, the community, and the culture of the projects. For GitHub projects, women have a higher pull request acceptance rate and more code review participation (except extra large projects) but longer code review interval time. The gender of the contributors has a mixed impact on Gerrit projects. So, a single strategy to mitigate bias in all projects might not be fruitful. We also investigated the effect of using gender-neutral profiles on the three aspects and provided recommendations based on our results.

Prospective newcomers in FOSS communities, project managers, and researchers may have some insights about the status of gender bias from our study and take appropriate steps to avoid biased occurrences and make a gender-inclusive environment.

\section*{Disclaimer}
\label{sec-disclaimer}
\revised{Our study primarily follows the proposed design published in the Registered Report~\cite{RR_ICSEME}, but there are a few deviations from the original study. We have listed those below:}

\begin{itemize}
    \item \revised{We have created both datasets based on more recent data. Originally we proposed to use the GHTorrent MySQL dump from June 2019. Instead, we have used the GHTorrent data dump from March 2021. Also, for the Gerrit dataset, we have updated Bosu et al.'s dataset to include all CRs completed till 30th April 2022. We have also identified bot accounts for exclusion. }
    \item \revised{A criterion for selecting GitHub projects was changed from at least 10 contributors to `at least 20 contributors', since we found that a random selection with lower thresholds increases selection space by orders of magnitude and returns projects that have low development activities.}

    \item \revised{To identify gender, we added a step, which is the identification of gender from the avatar.}

    \item \revised{There is a change in the computation of the gender-neutral profile attribute. We proposed in the study design that we would consider a profile gender-neutral if the gender of the user is not identifiable from the avatar, display name, full name, or by human ratings. But, we considered profiles as gender neutral if the gender of the profile owner is not identifiable from the avatar. We discarded inferring gender-neutralness from a name or using a human rater since it requires familiarity with the origin of the name. For example, most people from the Indian subcontinent may easily infer the name `Sayma' belongs to a woman. However, people from different origins may have difficulties. Moreover, even though a name-to-gender inference tool can reliably guess gender from a name, CR participants are less likely to use such tools during their collaborations. Therefore, to avoid such limitations, we considered only avatars as sources to identify gendered profiles, i.e., we marked a profile as GIP only if we can reliably identify its owner's gender using the associated avatar. }

    \item \revised{Moreover, we proposed that we would estimate the power of each independent variable using Wald Chi-Square statistics (Wald $\chi^2$).  During study execution, we used `Odds Ratio" (OR) for the logistic regression models and regression coefficients for linear regression models, instead.
    There are benefits from both ends in using one over the other.  Wald $\chi^2$ shows the influence of an independent variable over the trained model (i.e., how much explanatory power a model would lose if we remove this variable. On the other hand, OR shows how an independent variable's change influences the outcome variable. Specifically, for dichotomous variables, OR indicates how the odds of an outcome change if the factor is present in contrast to its absence. Since our variables of interest (i.e., gender and isGenderNeutral) are dichotomous, we think using OR has clear advantages in terms of interpreting the results.     
    Similarly, for linear regression models, the coefficient shows how the dependent variable changes, if an independent changes by one unit. Again, due to the dichotomous nature of our factors such as gender, the regression coefficient shows how the dependent changes if we change gender from `0' - `man' to `1' -`woman'. 
     Therefore, our modified choices have clear advantages over our prior selection in terms of interpretability.}
    
\end{itemize}

\section*{Data availability}
Our data mining and analysis scripts, and aggregated dataset are publicly available at \url{https://doi.org/10.5281/zenodo.760853}. Due to privacy concerns, we do not make the full dataset, which is more than 1TB publicly available. The authors would be happy to share the dataset with researchers upon contact.

\section*{Funding and Conflicts of interests/Competing interests.}
Work conducted by for this research is partially supported
by the US National Science Foundation under Grant No. 1850475. Any opinions,
findings, and conclusions or recommendations expressed in this material are those
of the author(s) and do not necessarily reflect the views of the National Science
Foundation. 

The authors have no competing interests to declare that are relevant to the content of this article.

\bibliographystyle{spmpsci}
\bibliography{references} 
\end{document}